\documentclass[lettersize,journal]{IEEEtran}
\usepackage{amsmath,amsfonts}
\usepackage{algorithmic}
\usepackage{algorithm}
\usepackage{array}
\usepackage[caption=false,font=normalsize,labelfont=sf,textfont=sf]{subfig}
\usepackage{textcomp}
\usepackage{stfloats}
\usepackage{url}
\usepackage{verbatim}
\usepackage{graphicx}
\usepackage{cite}
\usepackage{booktabs}

\usepackage{multirow}
\usepackage{makecell}
\usepackage{amsthm}
\theoremstyle{plain}

\theoremstyle{definition}

\theoremstyle{plain}

\theoremstyle{plain}

\begin{document}

\title{Asynchronous Multi-Class Traffic Management in Wide Area Networks}

\author{Hao Wu,~Jian Yan,~Linling Kuang
\thanks{ This work was supported in part by National Natural Science Foundation of China under Grant 62101300, in part by Shanghai Municipal Science and Technology Major Project under Grant 2018SHZDZX04, and in part by Tsinghua University Initiative Scientific Research Program under Grant 20221080009. (Corresponding author: Jian Yan.)

The authors are with the Department of Electronic Engineering, Tsinghua University, Beijing 100084, China, and also with the Beijing National Research Center for Information Science and Technology (BNRist), Tsinghua University, Beijing 100084, China.}
}
\maketitle

\begin{abstract}
The emergence of new applications brings multi-class traffic with diverse quality of service (QoS) requirements to wide area networks (WANs), motivating research in traffic engineering (TE). In recent years, novel centralized and hierarchical TE schemes have used heuristic or machine learning techniques to orchestrate resources in closed systems such as datacenter networks. However, these schemes suffer from long delivery delays and high control overhead when applied to general WANs. To provide low-delay services, this paper proposes an asynchronous multi-class traffic management (AMTM) scheme. We first establish an asynchronous TE paradigm in which distributed nodes locally perform low-complexity and low-delay traffic control based on link prices, and the TE server updates link prices to eliminate decision conflicts between edge nodes. By modeling the asynchronous TE paradigm as a control system with non-negligible control loop delay, we find that the traditional pricing strategy cannot simultaneously achieve a low packet loss rate and a low flow delivery delay. To address this issue, we propose a new pricing strategy based on the observations of virtual queues in intermediate nodes. We also present a system design and related algorithms that utilize a dynamic step size mechanism of link price update. Simulation results show that AMTM can effectively reduce the end-to-end flow delivery delay.
\end{abstract}

\section{Introduction}
With the emergence of new applications and the explosive growth of Internet traffic, ensuring high-quality service for multi-class traffic has become a major challenge for Internet service providers. In addition to investing in long-term infrastructure expansion, TE, a technology for improving QoS and network efficiency, is one way to tackle this issue.

In recent years, extensive research has been conducted on TE based on Software-Defined Networking (SDN)\cite{mendiola2016survey}. SDN  facilitates the deployment of online TE by enabling a centralized network architecture and automated configuration of the underlying devices. Compared to traditional TE approaches with distributed control logic\cite{AKYILDIZ20141,verma2016review,hanzo2007survey}, SDN-based TE achieves better performance in several metrics (e.g., link utilization\cite{10.1145/3230543.3230545}, network utility\cite{6678113}, and end-to-end (E2E) delay\cite{8277296}). The performance improvement of SDN-based TE can be attributed to three key factors. First, SDN-based TE can orchestrate network resources in a global network view. Second, flow-level control and automated control rule updates are enabled by SDN standards such as OpenFlow and P4\cite{10.1145/2656877.2656890}. In addition, complex heuristic algorithms or machine learning agents can be used for TE decision-making in the decoupled control plane, unconstrained by device capabilities. 

As the most popular branch of SDN-based TE, centralized TE schemes for private Inter-Datacenter Networks (IDNs) \cite{10.1145/3230543.3230545, RN33, 10.1145/2534169.2486012,7299623} typically utilize a periodic paradigm involving synchronous actions of network elements, including traffic matrix collection/prediction, resource allocation, and traffic control. This paradigm can avoid transient congestion caused by unsynchronized routing reconfiguration and service disruption caused by unsynchronized control rule updates. The success of these centralized TE schemes also depends on the controllability of traffic sources in IDNs. In IDNs, applications and servers can be controlled to predict traffic matrices for the next few minutes or even hours. Only some short real-time flows, typically less than 20$\%$ of the total traffic\cite{benson2010network}, are difficult to predict. Therefore, these approaches can achieve satisfactory performance by pre-allocating resources. 

However, many general WANs for communication may not possess the aforementioned characteristics as private IDNs. In such situations, the periodic paradigm may encounter the following problems. 

The first problem is the long E2E delay for flows. The network is unable to anticipate the demand of user-initiated flows in advance due to the unpredictable nature of user behavior. Furthermore, data mining results from previous WANs have revealed the volatility\cite{bai2015information} and unpredictability\cite{10.1145/2785956.2787472,10.1145/3487552.3487860} of particular traffic patterns. Consequently, traffic control rules based on traffic prediction may mismatch with the volatile traffic matrix, thereby resulting in performance penalties such as link overload and packet loss. An alternate strategy involves caching received flows at the network edge while awaiting periodic resource allocation. However, the waiting time, usually averaging 0.5 TE period, cannot be ignored for delay-sensitive flows due to the long TE period implemented to minimize update overhead\cite{8764418,7524330}. Although hierarchical TE architectures \cite{6678113,9110786} and learning-enabled TE algorithms \cite{xu2023teal} have reduced the algorithm runtime in each TE period to several seconds, it is still too long for delay-sensitive real-time flows.   

The second problem is the high control overhead during TE. Numerous flows with diverse QoS requirements exist within WANs, necessitating frequent interaction between the control plane and underlying devices to implement flow-level traffic control. During each TE period, a multitude of flow-level requests are uploaded to the control plane, and control instructions are frequently delivered to devices, resulting in significant control overhead.


To solve the aforementioned problems, we propose AMTM, a TE scheme that utilizes an asynchronous TE paradigm. In this paradigm, service brokers located in distributed edge nodes perform real-time traffic control based on link prices. To handle traffic mismatches caused by decision conflicts during distributed traffic control, AMTM buffers mismatched flows in intermediate nodes instead of dropping them, thereby increasing tolerance to traffic fluctuation. The TE server adjusts link prices, which impact the behaviors of service brokers, to eliminate decision conflicts and traffic mismatches. Moreover, a pricing strategy based on the observations of virtual queues is proposed to obtain a low packet loss rate and a low flow delivery delay. The main contributions of this paper are outlined as follows.

\hangafter 1 
\hangindent 1em 
\noindent 
$\bullet$ We establish an asynchronous TE paradigm and describe its working mechanisms, including flow-driven traffic control (FDTC) and network-initiated price update (NIPU) actions. FDTC is designed to ensure real-time flow delivery, while NIPU handles traffic mismatch.

\hangafter 1 
\hangindent 1em 
\noindent 
$\bullet$ We prove that the widely-used pricing strategy based on dual-decomposition cannot achieve satisfactory QoS in the asynchronous TE paradigm, resulting from the long control loop. Then, we propose a new pricing strategy based on the observations of virtual queues in intermediate nodes, which achieves better performance in packet loss and delay.  

\hangafter 1 
\hangindent 1em 
\noindent 
$\bullet$ We propose a system design of AMTM with a detailed description of the workflows and algorithms executed in edge nodes and the TE server. Simulated experiments based on a real network topology are performed to investigate the performance of AMTM in terms of network utility, flow delivery delay, and scalability.
\section{Related Work}
Early research on TE, which includes OSPF\cite{moy1997ospf}, ECMP, and QoS routing\cite{masip2006research} along with their variants, focused on distributed IP and Multi-Protocol Label Switching networks. QoS routing, designed to address constrained shortest routing problems\cite{guck2017unicast}, helps find the shortest path that satisfies specific QoS demands such as bandwidth and delay. However, these algorithms may result in suboptimal decisions due to the lack of a global network view\cite{AKYILDIZ20141}. 

The emergence of SDN enables TE based on a global network view\cite{7762818}. Over the past few years, many SDN-based centralized TE approaches have been proposed. For instance, Hong \emph{et al.}\cite{10.1145/3230543.3230545} applied centralized TE to Google's B4 network and managed to improve the average link utility by 2-3 times. Kandula \emph{et al.} \cite{RN33} proposed an online flow planning that improved the network utility in a Microsoft production WAN. Other approaches, such as SWAN \cite{10.1145/2534169.2486012} and MCTEQ \cite{7299623}, establish different centralized models to optimize throughput, network utility, and fairness. The aforementioned works employ various approximation heuristic algorithms to realize online TE for large-scale traffic.

Hierarchy control frameworks have been proposed in general WANs. Tomovic \emph{et al.}\cite{8764418} proposed a control framework that incorporates online routing and offline TE. Offline TE updates load balancing weights, and online routing provides a fast service based on these weights. The proposed framework effectively reduces the processing delay of traffic control for real-time service. Guck \emph{et al.}\cite{guck2016function} utilized network calculus to calculate the worst-case delay of "queue links", based on which an online admission control is executed. This method guarantees deterministic QoS for admitted flows. Ghosh \emph{et al.}\cite{6678113} decomposed the centralized TE problem into 2-tier and 3-tier semi-centralized problems, which reduces the computational complexity in the TE server. The authors \cite{9110786} have also proposed a 2-level algorithm to solve the service payoff maximization problem. The first-level algorithm is executed using parallel edge computing. 

Besides model-based TE, machine learning-based TE\cite{7557432} has also been studied in recent years. Geyer and Carle\cite{10.1145/3229607.3229610} utilized graph neural networks \cite{zhou2020graph} to solve distributed routing problems. They proposed an approach for training independent agents to generate distributed routing policies that can achieve a common goal (e.g., shortest path). Deep Reinforcement Learning (DRL) can handle complex states and actions in TE control decision-making. Using mode-free DRL, Xu \emph{et al.} \cite{8485853} factored E2E delay, which is not accurately modeled in traditional TE models, into network utility. By setting the network utility as the reward, DRL explores the optimal split ratio of each session under different throughputs and delays. Given that DRL can be time-consuming in large-scale networks, AuTO\cite{10.1145/3230543.3230551} is designed to provide quick decision-making for short flows. The peripheral system makes instant decisions locally for short flows, while the central system makes decisions for long flows. TEAL\cite{xu2023teal} utilizes graph neural networks to generate the structural feature vector of a network, which assistants the controller network to execute fast TE decision making. 

\section{System Model of TE}
In this section, we compare the periodic and asynchronous TE paradigms and introduce the mathematical model of TE. We use many key notations, as summarized in Table \ref{NTS}.
\begin{table}[ht]
\centering
\caption{List of Key Notations}
\begin{tabular}{p{1.5cm}<{\centering} p{6.5cm}}
\toprule
\multicolumn{1}{c}{\textbf{Symbol}} & \multicolumn{1}{c}{\textbf{Description}}\\ \midrule 
    $j,\mathcal{J}_t$ & Flow and flow set at time $t$\\
    $u_j(\cdot)$ & Utility function \\
    $p,\mathcal{P}$ & Path and candidate path set\\
    $x_{jp},x_{j}$ & Bandwidth allocated to flow $j$ on path $p$ and all paths  \\ 
    $\Theta_{jp}$ & Relation between flow $j$ and path $p$\\
    $\Phi_{pe}$ & Relation between path $p$ and link $e$\\
    $\Phi_{pe}^s$ & Relation between the $s^\text{th}$ link of path $p$ and link $e$\\
    $\mathcal{C}_e$ & Link capacity\\ 
    $\lambda_{e}$ & Link price ($\lambda_{et}$ represents its value at time $t$)\\
    $\mathcal{Q}_p^s$ & Virtual queue established for path $p$ in the $s^\text{th}$ link of $p$\\
    $r_{pt}^s$ & Traffic rate on the $s^\text{th}$ link of $p$ at time $t$\\
    $\alpha_{pt}^s,\beta_{pt}^s$ & Arrival and departure rates of virtual queue $\mathcal{Q}_p^s$\\
    $d_{pt}^s$ & Retention rate of virtual queue $\mathcal{Q}_p^s$\\
    $R_{pt}^s$ & Length of a virtual queue $\mathcal{Q}_p^s$\\
    $q_{et}$ & Queueing delay in the physical queue of link $e$\\
    $\mathcal{I}_{et}$ & Idle bandwidth of link $e$ at time $t$\\
    \bottomrule
  \end{tabular}
  \label{NTS}
\end{table}
\subsection{Overview of Periodic TE Paradigm}
\begin{figure}[th]
\centering
\includegraphics[height=1.7in]{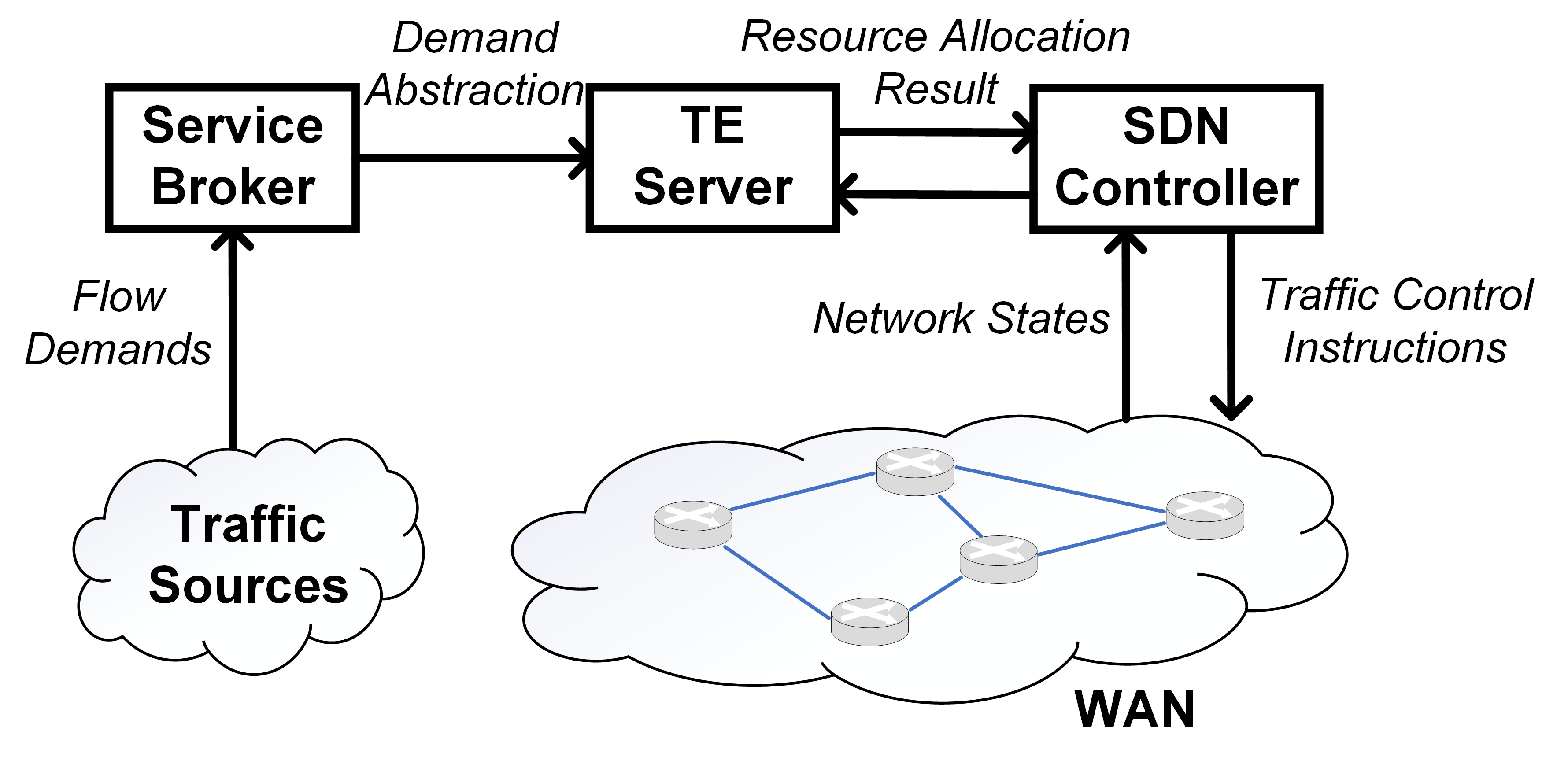}
\caption{The function elements and actions in the periodic TE paradigm.}
\label{fig1}
\end{figure}

As depicted in Fig. \ref{fig1}, the periodic TE paradigm, which is utilized by centralized TE schemes, follows a synchronized action loop during each TE period. Different function elements exhibit fixed behaviors in accordance with this action loop. Firstly, service brokers deployed in distributed nodes obtain flow demands from traffic sources. Once the collected data has been transformed into abstracted demands (e.g., utility functions, traffic type, QoS constraints), service brokers report them to the TE server. Subsequently, the TE server performs a fine-grained resource allocation for each flow based on the telemetry data of network states. The outcome is then announced to the SDN controller, which translates it into traffic control rules. Finally, the SDN controller rules configure the rules into underlying switches.

The above action loop is executed once in a TE period (e.g., five minutes). Flows that arrive after an action loop must wait for the next loop to receive resources, and the waiting time ranges from 0 to 1 TE period. The TE period cannot be too short in a WAN for the following three reasons. 

\hangafter 1 
\hangindent 1em 
\noindent
\romannumeral1) The round-trip time among traffic sources, the TE server, and underlying devices ranges from tens to hundreds of milliseconds. The total round-trip time of control messages can reach several seconds some scenarios (e.g., future 6G communications with space nodes included).

\hangafter 1 
\hangindent 1em 
\noindent
\romannumeral2) Fine-grained resource allocation algorithms are time-consuming because of the large variable scale. Hierarchical TE architectures \cite{6678113,9110786} and learning-enabled TE algorithms \cite{xu2023teal} reduced the algorithm runtime in each TE period to several seconds, which is still too long for delay-sensitive real-time flows.   

\hangafter 1 
\hangindent 1em 
\noindent
\romannumeral3) Frequent interaction between network elements is bandwidth-consuming because the abstracted demands and control instructions usually rely on in-band transmission. 

Thus, the TE periods in some existing schemes are set to several minutes. As a result, the low-latency requirement of some flows is not satisfied.  An alternative to provision real-time services is preallocating resources to flows one TE period earlier based on demand estimation. However, this approach suffers from traffic mismatch performance degradation when dealing with time-varying traffic (see Sec. VI).

The periodic TE paradigm also suffers a scalability problem. The TE server processes numerous flow queries and generates corresponding traffic control rules for them, making the processing capacity of the TE server the bottleneck of network scalability. 
\subsection{Asynchronous TE Paradigm}
\begin{figure*}[t]
\centering
\subfloat[]{\includegraphics[height=1.7in]{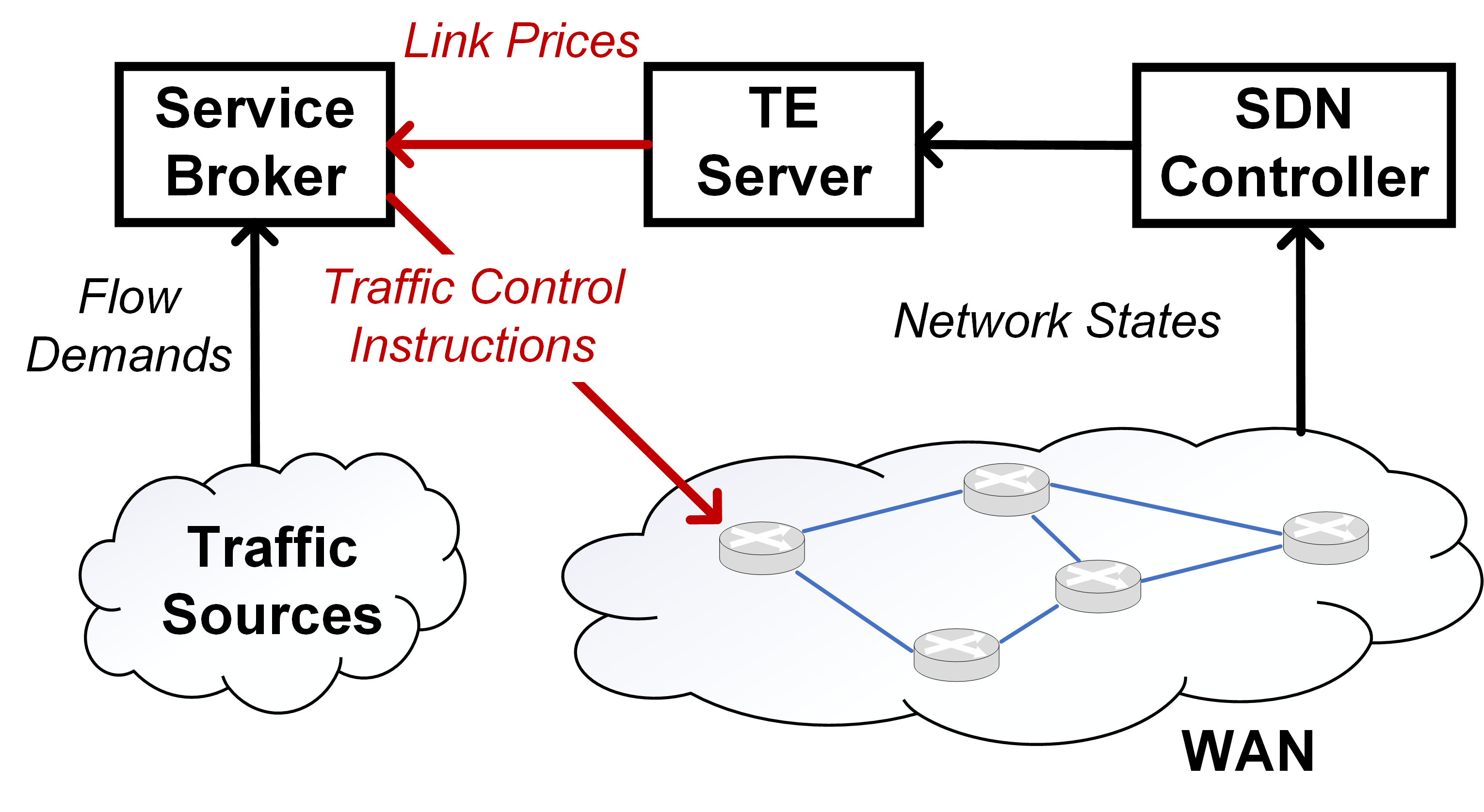}}\qquad \qquad
\subfloat[]{\includegraphics[height=2.0in]{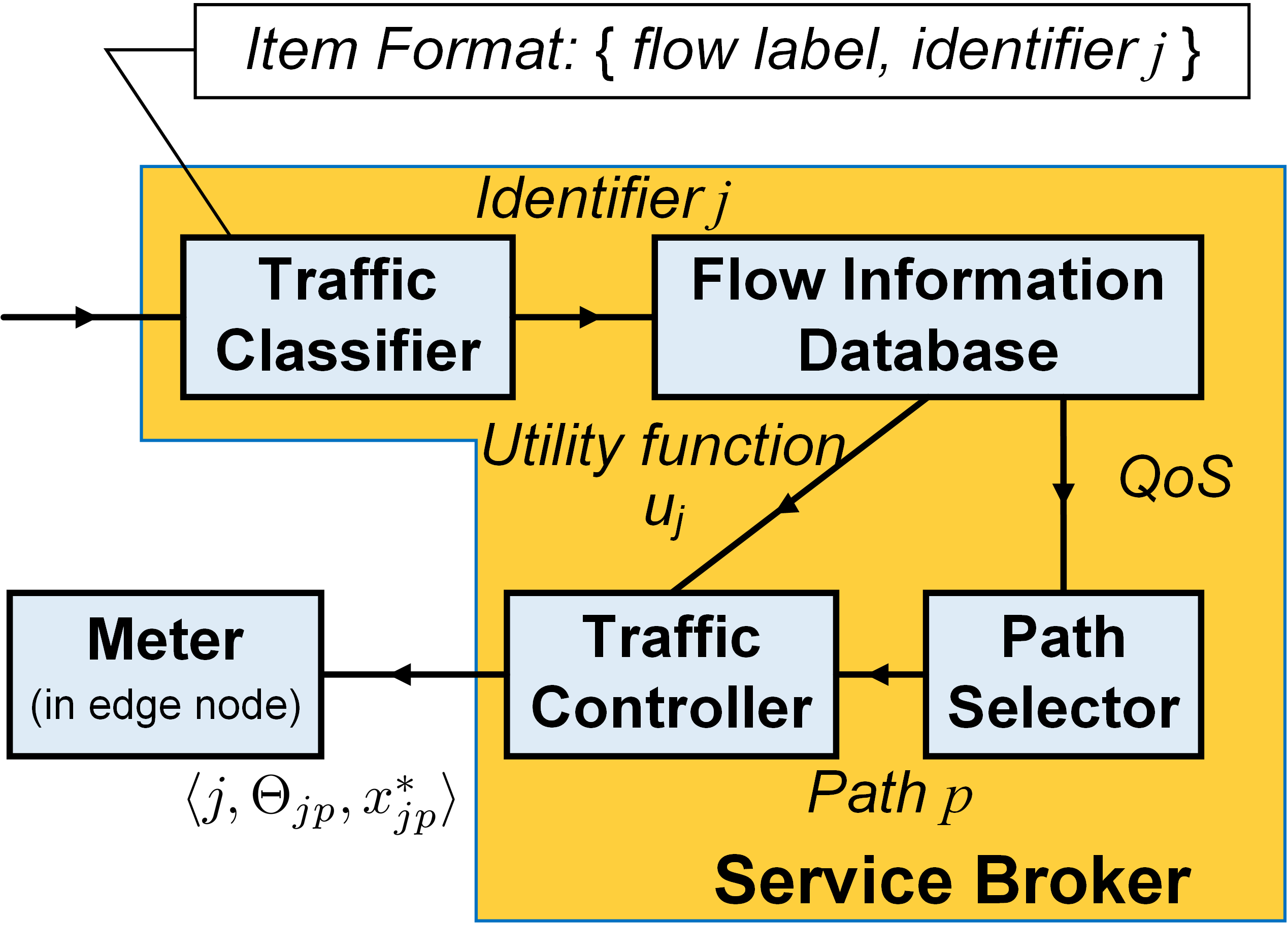}}
\caption{The asynchronous TE paradigm. (a) The function elements and actions. (b) The working process of FDCT in a service broker.}
\label{fig2}
\end{figure*}

In the paper, we establish an asynchronous TE paradigm without employing a strict action schedule. As depicted in Fig. \ref{fig2}(a), the differences between this paradigm and the above mentioned periodic TE paradigm are as follows. First, the decision-making of admission control and resource allocation, referred to as FDTC, is moved from the TE server to the service brokers at edge nodes\footnote{Flows enter the WAN through edge nodes. For example, edge and intermediate nodes are edge and core routers, respectively, in a backbone network.}. An FDTC action is instantly executed upon flow arrival, and the service broker makes a local decision with negligible control message round-trip time and decision making time. As a result, delay-sensitive flows experience immediate delivery. Second, the TE server detects and eliminates decision conflicts between independent service brokers by updating link prices, referred to as NIPU. 

\textbf{FDTC.} A FDTC action is activated with the arrival of each new flow. As illustrated in Fig. \ref{fig2}(b), flows first undergo a classification by a traffic classifier using its programmable flow table. The traffic classification policy, whose granularity can be session-level, user-level, service-level, or site-level, is determined by the network service provider. For instance, the classifier may utilize a user label (e.g., IP address) to recognize packets between a specific user pair and threat them as a distinct flow when applying a user-level policy. The service broker maintains the utility functions $u_j$ and QoS requirements of flows within a flow information database in accordance with prior knowledge and service-level agreements. For example, the QoS requirements of common applications such as industrial Internet, VoIP, and video streaming, are known in advance. Meanwhile, users can declare customized QoS parameters. The service broker queries the flow information database to achieve $u_j$ and the QoS requirements of a flow $j$. Afterwards, a path selector chooses one or more transmission paths from the candidate paths (denoted by a set $\mathcal{P}$) maintained by the network. For the sake of convenience, the routing outcome is represented by a matrix $\Theta$, with its element $\Theta_{jp}$ being 1 when a path $p\in \mathcal{P}$ is selected and 0 otherwise. Based on $u_j$, $\Theta_{jp}$, and the link prices provided by the TE server, a traffic controller calculates the allocated bandwidth $x_{jp}^*$ to the flow on each selected path. Lastly, the service broker sends a control instruction accompanied by a tuple $\langle j,\Theta_{jp},x_{jp}^* \rangle$ to the  meter of the edge node, which adds meter rules to prevent the rate of flow $j$ on each path $p$ from exceeding $x_{jp}^*$.

\textbf{NIPU.} The service brokers execute FDTC actions independently without negotiation. Then, decision conflicts on traffic control may occur, causing link overload or underload. In this context, a NIPU action is activated by the TE server at a fixed frequency, or when a preset condition (e.g., the detection of congestion, queue overflow, and excessive delay) is satisfied. Firstly, the SDN controller obtains the telemetry result of network states from intermediate nodes and uploaded it to the TE server. Following this, the TE server calculates new link prices and transmits them to service brokers. Adjusting link prices changes the traffic control behaviors of distributed service brokers, resulting in the transition of network state. Then, undesired events, such as link congestion, packet loss, and long queueing delay, caused by decision conflicts of service brokers are eliminated. Designing a link pricing strategy in a time-varying system is a challenge, and our solution is proposed in Sec. IV. 
\subsection{Utility Optimization Model}
Besides ensuring a satisfactory QoS, TE designers also strive to enhance system efficiency, which can be evaluated by network utility. A WAN is usually represented by a directed graph $\mathcal{G}=(\mathcal{N},\mathcal{E})$ with a set $\mathcal{N}$ of switch nodes and a set $\mathcal{E}$ of links. When receiving a packet, node $n\in \mathcal{N}$ forwards it to an egress link or places it into its egress queue. The maximum link rates are denoted by $\mathcal{C}_e$, and the network maintains a set $\mathcal{P}$ of candidate paths for flows. The relation between paths and links is represented by matrix $\Phi$. When path $p$ contains link $e$, $\Phi_{pe}=1$; otherwise, $\Phi_{pe}=0$. Then, the utility optimization problem can be expressed as
\begin{equation}
\begin{aligned}
\text{maximize  } &\mathcal{U}=\sum_{j\in \mathcal{J}_t} u_j(x_j) && (\ref{ONU}.a)\\
\text{subject to } &\sum_{j\in \mathcal{J}_t}\sum_{p\in \mathcal{P}} \Theta_{jp} \Phi_{pe} x_{jp} \le  \mathcal{C}_e,\forall e && (\ref{ONU}.b)\\
&\sum_{p\in\mathcal{P}}x_{jp} = x_j,\forall j && (\ref{ONU}.c)\\
\text{variables  } &x_j,x_{jp}\ge 0,\forall j,p && (\ref{ONU}.d)
\end{aligned}
\label{ONU}
\end{equation}
where $\mathcal{U}$ is the network utility and $\mathcal{J}_t$ contains all the ongoing flows at $t$. Constraint (\ref{ONU}.b) ensures that the allocated bandwidth on each link does not exceed its capacity. For the sake of convenience, we represent problem (\ref{ONU}) by $\mathcal{U}(\mathcal{J}_t,\mathbf{u}_t,\Theta,\Phi,\mathcal{C})$, where $\mathbf{u}_t=\{u_j|j\in\mathcal{J}_t\}$ and $\mathcal{C}=\{\mathcal{C}_e|e\in\mathcal{E}\}$. 

Specific utility function designs have various practical implications. Table \ref{Utility} shows some typical utility functions and their respective practical implications mentioned in previous studies. For the sake of convenience, we use the same notations to express them. A common feature of these functions is that they are concave functions $u_j(x_j)$ of flow transmission rate $x_j$. The concavity of $u_j(x_j)$ is reasonable since users achieve diminishing returns from received bandwidth, especially after adequate bandwidth is allocated. This paper provides a solution to the general utility maximization problem and does not make assumptions about the expression of the utility function aside from diminishing marginal. 
\begin{table}[htb]
\caption{Utility functions for different TE objective}
\label{Utility}
\centering
\begin{tabular}{p{5cm} p{2.8cm}<{\centering}}
\toprule
\multicolumn{1}{c}{\textbf{Utility function}} & \multicolumn{1}{c}{\textbf{Implication}}\\ \midrule
$u_j(x_j)=w_k\log(1+x_j)$, where $w_k$ is the weight of each traffic class. & Proportional fairness among flow groups \cite{7299623}\\
 \hline
$u_j(x_j)=w_k[a_kf_k(x_j)-b_kg_k(d_j)]$, where $f_k(x_j)$ and $g_k(d_j)$ are rate-dependent and delay-dependent functions, respectively. Parameters $a_k$ and $b_k$ have specific values for different traffic classes. & E2E delay and throughput \cite{8485853}\\ \hline
 $u_j(x_j)= \frac{w_j}{1-\alpha}x_j^{1-\alpha}$, where $w_j$ is the weight of a flow. & $\alpha$-fairness \cite{srikant2004mathematics}\\\hline
$u_j(x_j)=w_jD_jF_k(x_j(t))$, where $D_j$ is the data volume. Function $F_k(\cdot)$ measures the service consistency based on the allocated bandwidth $x_j(t)$ at several periods. & Service consistency and throughput \cite{9110786}\\
\bottomrule
\end{tabular}
\end{table}
\section{Link Pricing Strategy of AMTM}
In this section, we focus on link pricing strategies. First, we show that the traditional link pricing strategy based on dual-decomposition surfers unsatisfactory QoS in the asynchronous TE paradigm. Then, we propose our solution. 
\subsection{Drawbacks of the link pricing strategy based on dual decomposition}
Link pricing can be used to solve the utility optimization problem. A pricing strategy based on the widely-used dual-decomposition\cite{1664999,huang2013wireless} is as follows. 
\begin{equation}
\begin{aligned}
&x_{jp}^*\leftarrow \mathop{\arg\max}_{x_{jp}\ge 0}u_j(\sum_{p\in\mathcal{P}}x_{jp})-\sum_{e\in \mathcal{E}}\lambda_e\sum_{p\in\mathcal{P}}x_{jp}\Theta_{jp}\Phi_{pe},\\
&\lambda_e \leftarrow [\lambda_e + \mu \dot{\lambda}_{e}]^+,\quad \dot{\lambda}_{e} = \sum_{j\in \mathcal{J}_t}\sum_{p\in\mathcal{P}}x_{jp}^*\Theta_{jp}\Phi_{pe}-\mathcal{C}_e,
\end{aligned}
\label{TraditionPricing}
\end{equation}
where $[\cdot]^+$ denotes the projection onto the nonnegative orthant. The first line determines the bandwidth $x_{jp}^*$ allocated to each flow $j$ on each path $p$, while the second line provides the iteration strategy of price $\lambda_e$ on each link $e$. The derivation details and convergence proof of this pricing strategy are provided in Appendix A. Next, we will demonstrate that employing this pricing strategy directly in the asynchronous TE paradigm leads to packet loss and long queueing delays.
\begin{figure*}[t]
\centering
\includegraphics[width=2\columnwidth]{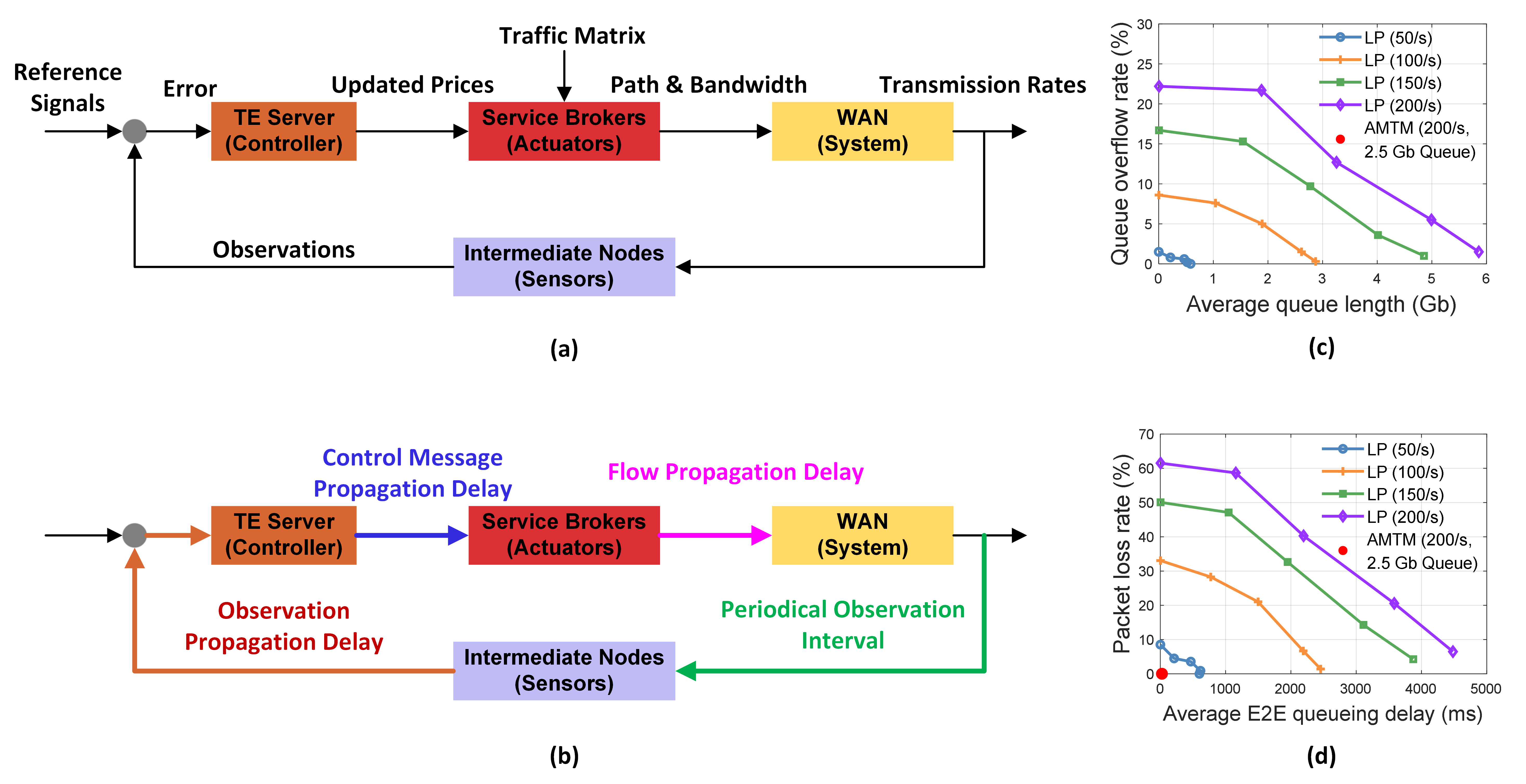}
\caption{Performance of the asynchronous TE paradigm employing the traditional LP solution. (a) Control model. (b) Control delay. (c) Link states. (d) QoS experienced by flows. The markers from left to right in each LP curve correspond to the queue depths of 0, 2.5 Gb, 5 Gb, 10 Gb, and 15 Gb. The flow arrival intensities are shown in parentheses. More simulation details are described in Sect. \ref{NumericalResult}.}
\label{figTrad}
\end{figure*}

As depicted in Fig. \ref{figTrad}(a), the network using the asynchronous TE paradigm can be conceptualized as a control system, where the TE server, service brokers, WAN, and intermediate nodes can be considered as the controller, actuators, system, and sensors, respectively. The TE server distributes updated prices to service brokers, and service brokers, based on the time-varying traffic matrix, determine the bandwidth allocation $x_{jp}^*$ on the candidate paths. The allocation results are configured to the data plane of the WAN. As traffic enters the WAN, intermediate nodes generate network state observations such as link overload or underload. The error between the reference signals (e.g., QoS demands and reference values of network states) and the observations is used by the TE server in the price update. 

For the sake of convenience, we add a subscript $t$ to link prices to represent time. For example, $\lambda_{et}$ represents the link price at time $t$, and $\dot{\lambda}_{et}=\sum_{j\in \mathcal{J}_t}\sum_{p\in\mathcal{P}}x_{jp}^*\Theta_{jp}\Phi_{pe}-\mathcal{C}_e$ represents the gradient of $\lambda_{et}$ based on the pricing decision at time $t$. If the control system has a negligible delay during the control loop, link prices are guaranteed to converge to the current optimal prices $\lambda_{et}^*$ before the traffic matrix changes, as depicted in Table \ref{Delay_Convergence}. Consequently, the WAN achieves maximum network utility without encountering link congestion. However, a non-negligible control delay, consisting of four terms as depicted in Fig. \ref{figTrad}(b), exists in real WANs. Control messages experience propagation delay from the TE server to service brokers, and observations experience propagation delay from intermediate nodes to the TE server. Intermediate nodes can sense network state transitions and generate corresponding observations after flow propagation delay from traffic sources to intermediate nodes. These three propagation delay terms typically range from tens to hundreds of milliseconds in WANs. Additionally, observations are generated and uploaded periodically, resulting in a periodical observation interval within the control loop.
\begin{table}[htb]
\caption{Convergence of the link pricing strategy}
\centering
\begin{tabular}{p{2cm}<{\centering} p{3.5cm}<{\centering} p{2cm}<{\centering}}
\toprule
\multicolumn{1}{c}{\textbf{Control Loop}} & \multicolumn{1}{c}{\textbf{Pricing Iteration}} & \multicolumn{1}{c}{\textbf{Convergence}}\\ \midrule
negligible delay & $\lambda_{et} \leftarrow [\lambda_{et} + \mu \dot{\lambda}_{et}]^+$ & Guaranteed \\
delay $d$ & $\lambda_{e(t+d)} \leftarrow [\lambda_{et} + \mu \dot{\lambda}_{et}]^+$ & Usually not convergent
\\ \bottomrule
\end{tabular}
\label{Delay_Convergence}
\end{table}

We denote the total delay by $d$. The observations received by the TE server at time $t+d$ are actually determined by the pricing decision made at time $t$. Under time-varying traffic matrices, the values of $\dot{\lambda}_{e(t+d)}$ and $\dot{\lambda}_{et}$ are usually different. As a result, the actual link prices $\lambda_{e(t+d)}=\int_0^t \mu\dot{\lambda}_{ex} dx$ usually do not converge to the optimal link prices $\lambda_{e(t+d)}^*$, resulting in link overloads or underloads. In this case, the network suffers from a high packet loss rate if the intermediate nodes have shallow queues. Fig. \ref{figTrad}(c)(d) shows the performance curve of the link pricing (LP) solution in (\ref{TraditionPricing}). In each LP curve, the leftmost marker indicates the performance when the intermediate nodes have zero queue depth. The results show a high queue overflow rate and packet loss rate at different traffic arrival intensities, leading to user experience degradation and frequent retransmission.

Deep queues can be added to intermediate nodes to buffer overloaded traffic and reduce packet loss. If the network uses the LP solution in (\ref{TraditionPricing}), queue overflow and packet loss may still occur. As we increase the queue depth (i.e., move to the right along the performance curves in Fig. \ref{figTrad}(c)), queue overflow rates decrease at the cost of incremental queueing time in each node. Consequently, flows experience lower packet loss rates at the expense of an incremental flow queueing delay, as shown in Fig. \ref{figTrad}(d). In summary, the LP solution cannot simultaneously achieve a low packet loss rate and low flow queueing delay.

To address this issue, our method, AMTM, incorporates queueing states into its pricing strategy. Fig. \ref{figTrad}(c)(d) also presents the performance of AMTM under a flow arrival rate of 200/s. Using a 2.5 Gb-depth queue for each link, AMTM achieves a 0 queue overflow rate, 0 packet loss rate, 3.83 ms average queueing time per node and 12.58 ms average flow queueing delay, which is significantly better than the LP solution. The proposed pricing strategy is described in the following section.
\subsection{Link pricing strategy based on virtual queues in intermediate nodes}
\begin{figure}
\centering
\includegraphics[width=2.8 in]{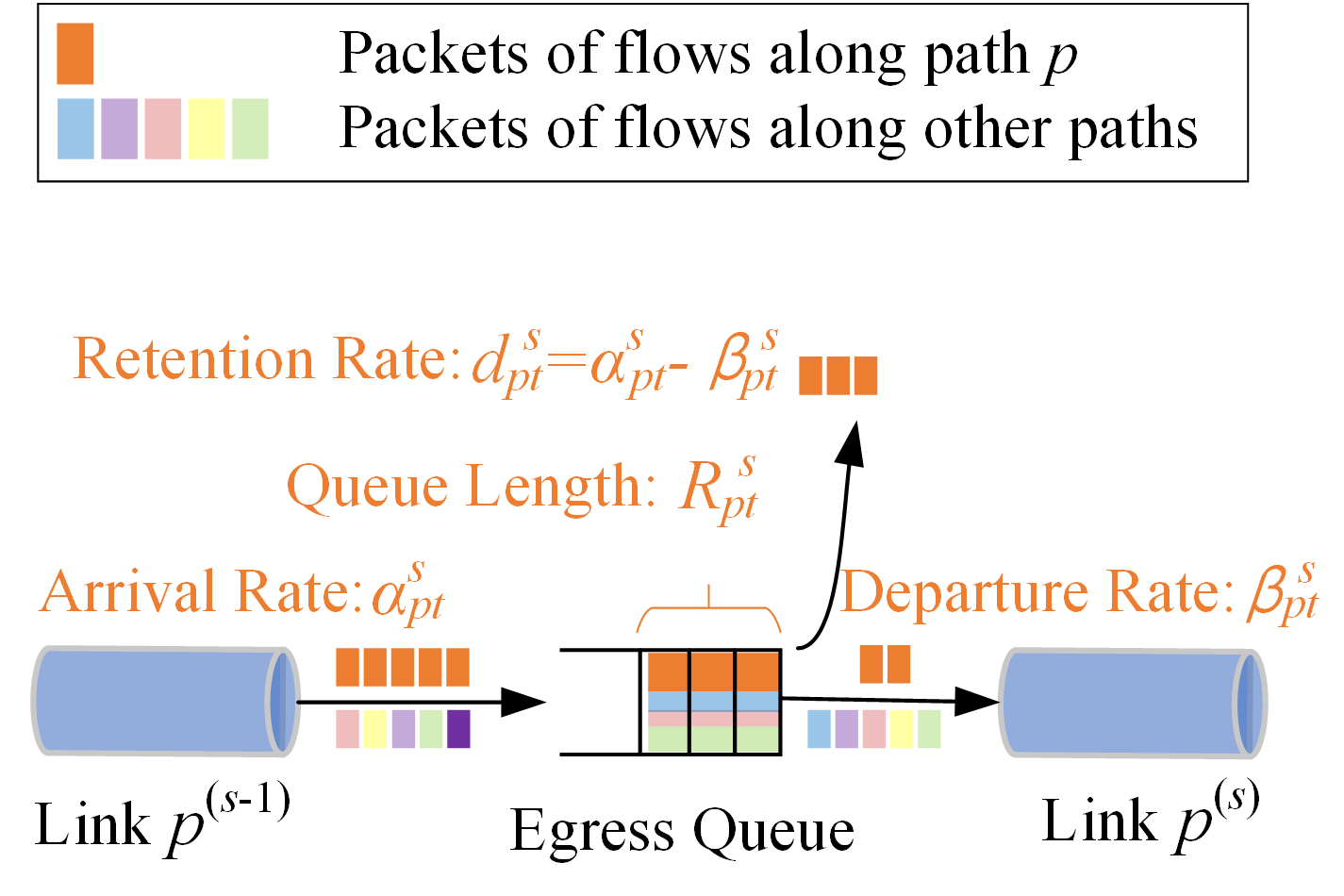}
\caption{The observations from virtual queue $\mathcal{Q}_p^s$.}
\label{fig13}
\end{figure}

To handle link overload, nodes can utilize a finite-length queue for each egress link, allowing packet buffering during link overload. The physical queue of an egress link is divided into virtual queues corresponding to all the candidate paths through the link. For convenience, we use $p^{(s)},s=1,...,|p|$ to represent the $s^{th}$ link in path $p$, whose length is denoted by $|p|$. We denote the relation between $p^{(s)}$ and $e$ by $\Phi_{pe}^s$. If $p^{(s)}$ is link $e$, $\Phi_{pe}^s=1$; otherwise, $\Phi_{pe}^s=0$. A virtual queue, denoted by $\mathcal{Q}_p^s$, presents in the intersection node of links $p^{(s-1)}$ and $p^{(s)}$ to buffer the packets transmitted along path $p$. A group of virtual queues $\{\mathcal{Q}_p^s|\Phi_{pe}^s=1\}$ share the physical egress queue of link $e$. The arrival rate $\alpha_{pt}^s$ and departure rate $\beta_{pt}^s$ of $\mathcal{Q}_p^s$ at time $t$ is observed and recorded, and their difference is referred to as retention rate $d_{pt}^s$. Meanwhile, the length $R_{pt}^s$ of $\mathcal{Q}_p^s$ and the idle bandwidth $\mathcal{I}_{et}$ of the egress link $e$ are also observed and recorded. As shown in Fig. \ref{fig13}, these observations are collected for link price update. The mathematical expressions for these observations are as follows.

When a flow $j$ is forwarded through $p$, its real transmission rate in $p^{(s)}$ may be lower than its rate $x_{jp}^*$ admitted by service brokers. if congestion occurs in $p^{(s)}$. In this situation, the packets are forwarded with an average probability $f_{pt}^s\in [0,1]$ in $p^{(s)}$ at $t$. We get $f_{pt}^s=f_{\hat{p}t}^{\hat{s}},\forall\hat{p}^{(\hat{s})}=p^{(s)}$ if egress queues treat packets along different paths fairly. For flows that select path $p$, we denote their total transmission rate in link $p^{(s)}$ by 
\begin{equation}
r_{pt}^s = \sum_{j\in \mathcal{J}_t}x_{jp}^*\Theta_{jp}\prod_{k=1}^{s}  f_{pt}^k, \quad s \in [1,|p|],
\label{rpt}
\end{equation}
which equals the departure rate $\beta_{pt}^s$ of $\mathcal{Q}_p^s$. For the sake of convenience, we let $r_{pt}^0 = \sum_{j\in \mathcal{J}_t}x_{jp}^*\Theta_{jp}$. Then, the value of arrival rate $\alpha_{pt}^s = r_{pt}^{s-1}$. In this situation, the node can calculate the retention rate $d_{pt}^{s}$ of packets along path $p$ in $\mathcal{Q}_p^s$ as
\begin{equation}
d_{pt}^{s}= \alpha_{pt}^s - \beta_{pt}^s = r_{pt}^{s-1} - r_{pt}^s = r_{pt}^{s}\frac{1-f_{pt}^{s}}{f_{pt}^{s}}.
\label{d}
\end{equation}
Moreover, link $e$ may have idle bandwidth when the total transmission rate in it is less than the link capacity. We can express the idle bandwidth $\mathcal{I}_{et}$ of link $e$ at $t$ as
\begin{equation}
\mathcal{I}_{et}=\max(\mathcal{C}_e-\sum_{p\in\mathcal{P}}\sum_{s=1}^{|p|}\Phi_{pe}^s r_{pt}^s,0).
\label{idle}
\end{equation}
The length $R_{pt}^s$ of $\mathcal{Q}_p^s$ can be directly obtained by the node. These observations have the following relation under a specific queueing mechanism. 

When the packets arrive at $Q_p^s$ in a random order and $Q_p^s$ utilizes a commonly-used first-in-first-out mechanism, the arrival rate $\alpha_{pt}^s$ and departure rate $\beta_{pt}^s$ of $Q_p^s$ are
\begin{equation}
\begin{aligned}
\alpha_{pt}^s&=r_{pt}^{s\text{-1}},\\
\beta_{pt}^s&=r_{pt}^s=\left\lbrace\begin{aligned}
\frac{(\mathcal{C}_e-\mathcal{I}_{et})R_{pt}^s}{\sum_{\hat{p},\hat{s}}\Phi_{\hat{p}e}^{\hat{s}} R_{\hat{p}t}^{\hat{s}}},&\text{ if $\sum_{\hat{p},\hat{s}}\Phi_{\hat{p}e}^{\hat{s}}R_{\hat{p}t}^{\hat{s}} \neq 0$}\\
\frac{(\mathcal{C}_e-\mathcal{I}_{et})r_{pt}^{s\text{-1}}}{\sum_{\hat{p},\hat{s}}\Phi_{\hat{p}e}^{\hat{s}}r_{\hat{p}t}^{\hat{s}\text{-1}}},&\text{ else if $r_{pt}^{s\text{-1}} \neq 0$}\\
0,&\text{ otherwise}
\end{aligned}\right.\end{aligned}
\label{alpha_beta}
\end{equation}
where $R_{pt}^s$ is the length of virtual queue $Q_{pt}^s$ as
\begin{equation}
R_{pt}^s = \max_{t'}\int_{t'}^t (\alpha_{p\tau}^s-\beta_{p\tau}^s) d\tau.
\label{R}
\end{equation} 

Based on the aforementioned observations on virtual queues, we propose the following iterative strategy.  
\\\textbf{Theorem 1:} \emph{If the optimal prices $\lambda^*_{et},\forall e\in \mathcal{E}$ of the network utility maximization problem $\mathcal{U}(\mathcal{J}_t,\mathbf{u}_t,\Theta,\Phi,\mathcal{C})$ fluctuate at a slower rate in comparison to the iteration convergence process, the asynchronous TE paradigm can make the network converge to the maximum network utility by employing the following strategies.}
\begin{equation}
\begin{aligned}
& x_{jp}^* \leftarrow \mathop{\arg\max}_{x_{jp}\ge 0}u_j(\sum_{p\in\mathcal{P}}x_{jp})-\sum_{e\in \mathcal{E}}\lambda_{et}\sum_{p\in\mathcal{P}}x_{jp}\Theta_{jp}\Phi_{pe},\forall j,p\\
& \lambda_{e(t+d)} \leftarrow \left[\lambda_{et} + n\dot{\lambda}_{et}^{(1)} + \mu \dot{\lambda}_{et}^{(2)}\right]^+,\quad\forall e\in \mathcal{E}\\
& \dot{\lambda}_{et}^{(1)} = \sum_{p\in\mathcal{P}}\sum_{s=1}^{|p|}\sum_{k=1}^{s}\Phi_{pe}^s R_{pt}^k,\quad\forall e\in \mathcal{E}\\
& \dot{\lambda}_{et}^{(2)} = \sum_{p\in\mathcal{P}}\sum_{s=1}^{|p|}\Phi_{pe}^s\mathcal{O}_{pt}^{s}-\mathcal{I}_{et}, \quad\forall e\in \mathcal{E}
\end{aligned}
\label{Corollary1}
\end{equation} 
\emph{where $n$ and $\mu$ represent step sizes, and $\mathcal{O}_{pt}^{s}$ represents the overload rate of link $p^{(s)}$ as}
\begin{equation}
\mathcal{O}_{pt}^{s} = \sum_{k=1}^{s}d_{pt}^{k}=r_{pt}^{0}-r_{pt}^{s}.
\label{Overload} 
\end{equation}
\emph{Specifically, the first equation in (\ref{Corollary1}) is utilized by service broker to determine the bandwidth $x_{jp}^*$ allocated to flow $j$ on path $p$, while the second equation in (\ref{Corollary1}) is utilized by the TE server to update link prices.}

\begin{proof}
We represent the optimal dual solution of the following problem by iteration target $\hat{\lambda}_{et}$.
\begin{equation}
\begin{aligned}
&\text{maximize  } \mathcal{U}=\sum_{j\in\mathcal{J}_t} u_j(\sum_{p\in\mathcal{P}}x_{jp})\\
&\text{s.t.} \sum_{j\in\mathcal{J}_t}\sum_{p\in \mathcal{P}} \Theta_{jp} \Phi_{pe} x_{jp}\le  \mathcal{C}'_{e},\forall e \\
&\text{variables  } x_{jp}\ge 0,\forall j,p
\end{aligned}
\label{ONU2}
\end{equation}
where $\mathcal{C}'_{e} = \mathcal{C}_e - \frac{n}{\mu}\dot{\lambda}_{et}^{(1)}$. The square error between $\lambda_{et}$ and $\hat{\lambda}_{et}$ can be measured by 
\begin{equation}
\mathcal{F}_t=\frac{1}{2}\sum_{e\in\mathcal{E}}(\lambda_{et}-\hat{\lambda}_{et})^2.
\end{equation} 
When the step sizes are small enough, we get
\begin{equation}
\Delta\mathcal{F}_t = \sum_{e\in\mathcal{E}}(\lambda_{et}-\hat{\lambda}_{et})\Delta\lambda_{et}-\sum_{e\in\mathcal{E}}(\lambda_{et}-\hat{\lambda}_{et})\Delta\hat{\lambda}_{et},
\label{Ft}
\end{equation}
where the symbol $\Delta$ represents increment after delay $d$. For example, $\Delta\mathcal{F}_t = \mathcal{F}_{t+d}-\mathcal{F}_t$. 
Then, we analyze the values of the first and second terms.

\textbf{The first term in (\ref{Ft}).} According to the expression of $\mathcal{O}_{pt}^s$ in (\ref{Overload}), the following term in the iteration increment of link prices satisfies
\begin{equation}
\dot{\lambda}_{et}^{(2)}= \sum_{p\in\mathcal{P}}\sum_{s=1}^{|p|}\Phi_{pe}^s r_{pt}^0 - \sum_{p\in\mathcal{P}}\sum_{s=1}^{|p|}\Phi_{pe}^s r_{pt}^s-\mathcal{I}_{et}.
\label{ProofT1}
\end{equation}
When link $e$ is overloaded, the second and third terms equal $\mathcal{C}_e$ and $0$, respectively; otherwise, the sum of the second and third terms are $\mathcal{C}_e$ according (\ref{idle}). In this context, we plug the expression of $r_{pt}^0$ into equation (\ref{ProofT1}) and get
\begin{equation}
\dot{\lambda}_{et}^{(2)} = \sum_{j\in\mathcal{J}_t}\sum_{p\in\mathcal{P}}\Phi_{pe}x_{jp}^*\Theta_{jp}-\mathcal{C}_e.
\label{ProofT1_2}
\end{equation}
According to (\ref{ProofT1_2}) and the expression of $\mathcal{C}'_{e}$, the price iteration in (\ref{Corollary1}) satisfies
\begin{equation}
\lambda_{e(t+d)}\leftarrow\left[\lambda_{et} + \mu\left(\sum_{j\in\mathcal{J}_t}\sum_{p\in\mathcal{P}}\Phi_{pe}x_{jp}^*\Theta_{jp}-\mathcal{C}'_{e}\right)\right]^+.
\end{equation}
We first analyze the case that the value in $[\cdot]^+$ is positive. Similar to (\ref{varphi}) in the proof of Lemma A.1, the objective function $\varphi(\lambda)$ in the Lagrange dual problem of (\ref{ONU2}) satisfies
\begin{equation}
\frac{\partial\varphi(\lambda)}{\partial\lambda_{et}} = \mathcal{C}'_e - \sum_{j\in\mathcal{J}_t}\sum_{p\in\mathcal{P}}\Phi_{pe}x_{jp}^*\Theta_{jp}=-\frac{1}{\mu}\Delta\lambda_{et}.
\end{equation} 
Meanwhile, $\varphi(\lambda)$ is a concave function\footnote{This conclusion can be proved in many ways. If you are interested in the proof, see https://math.stackexchange.com/questions/1374399/why-is-the-lagrange-dual-function-concave} of $\lambda_{et}$, and $\hat{\lambda}_{e}$ is the optimal solution of (\ref{ONU2}) corresponding to the minimum value of $\varphi(\lambda)$. Thus, we get
\begin{equation}
\begin{aligned}
&\sum_{e\in\mathcal{E}}(\lambda_{et}-\hat{\lambda}_{et})\Delta\lambda_{et} = - \mu \sum_{e\in\mathcal{E}}(\lambda_{et}-\hat{\lambda}_{et})\frac{\partial\varphi}{\partial\lambda_{et}}\\
&\le -\mu(\varphi(\lambda)|_{\lambda_e=\lambda_{et}}-\varphi(\lambda)|_{\lambda_e=\hat{\lambda}_{et}})\le 0,
\end{aligned}
\end{equation}
which means the first term in (\ref{Ft}) is non-positive. 

When the value in $[\cdot]^+$ is negative, we get $0\ge \Delta\lambda_{et} \ge -\mu(\mathcal{C}'_e-\sum_{j\in\mathcal{J}_t}\sum_{p\in\mathcal{P}}\Phi_{pe}x_{jp}^*\Theta_{jp}) = -\mu\frac{\partial\varphi(\lambda)}{\partial\lambda_{et}}$. The concavity of $\varphi(\lambda)$ indicates $-\mu\frac{\partial\varphi(\lambda)}{\partial\lambda_{et}}(\lambda_{et}-\hat{\lambda}_{et})\le 0$. Then, $\lambda_{et}-\hat{\lambda}_{et}\ge 0$ and $(\lambda_{et}-\hat{\lambda}_{et})\Delta\lambda_{et}\le 0$ hold for any $e\in\mathcal{E}$, which means the first term in (\ref{Ft}) is non-positive. 

\textbf{The second term in (\ref{Ft}).} The difference between problems (\ref{ONU}) and (\ref{ONU2}) is the values $\mathcal{C}_e$ and $\mathcal{C}'_e$. If the values of $R_{pt}^k$ converge to 0, then $\dot{\lambda}_{et}^{(1)}$ and $\mathcal{C}'_e$ converge to 0 and $\mathcal{C}_e$, respectively. In this situation, the optimal prices $\hat{\lambda}_{et}$ of (\ref{ONU2}) converge to the optimal prices of (\ref{ONU}). Then, we prove that $R_{pt}^s$ is decreasing until 0 in the iteration process.

According to the relation in (\ref{alpha_beta}), we get that the departure rate $\beta_{pt}^s$ will finally converge to a lower value if we reduce the arrival rate $\alpha_{pt}^s$ when $R_{pt}^s\neq 0$. Fig. \ref{fig3} shows the time-varying processes of $\alpha_{pt}^s$ and $\beta_{pt}^s$ in different cases. In case (a), $\frac{\text{d}R_{pt}^s}{\text{d}t}=\alpha_{pt}^s-\beta_{pt}^s<0$ at the beginning. Then, $\beta_{pt}^s = \frac{(\mathcal{C}_e-\mathcal{I}_{et})R_{pt}^s}{\sum_{\hat{p},\hat{s}}\Phi_{\hat{p}e}^{\hat{s}} R_{\hat{p}t}^{\hat{s}}}$ decreases as $R_{pt}^s$ decreases until $\alpha_{pt}^s=\beta_{pt}^s$. In case (b), $\frac{dR_{pt}^s}{dt}=\alpha_{pt}^s-\beta_{pt}^s>0$ at the beginning. Then, $\beta_{pt}^s$ increases until it equals $\alpha_{pt}^s$ at a specific time $t'$, after which $\beta_{pt}^s$ decreases as in case (a). The analysis result reveals that $\beta_{pt}^s$ and $R_{pt}^s$ are guaranteed to decrease after an inflection time $t'$ if $\alpha_{pt}^s$ keeps decreasing. 
\begin{figure}[tbh]
\centering
\subfloat[]{\includegraphics[height=1.2in]{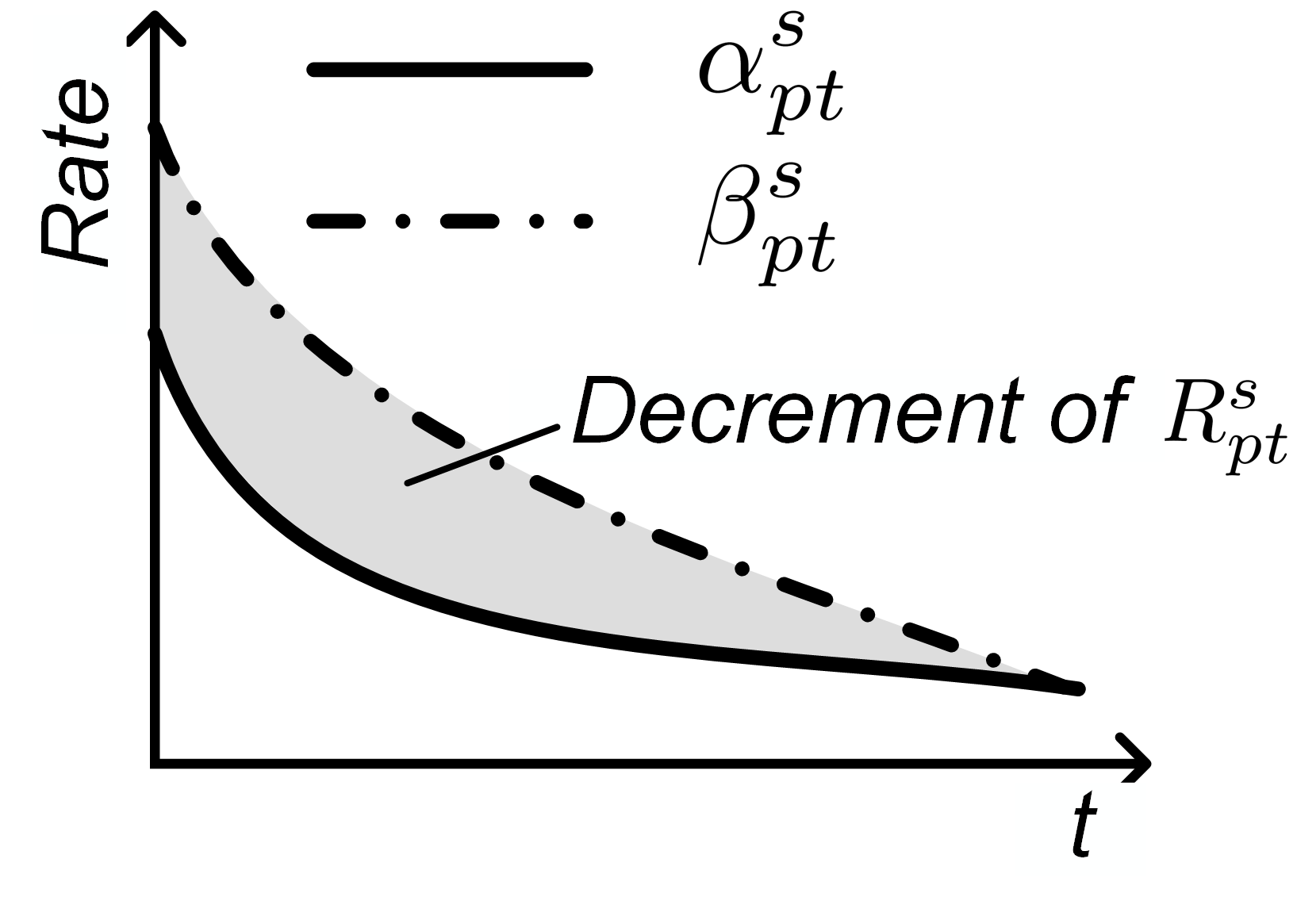}}
\subfloat[]{\includegraphics[height=1.2in]{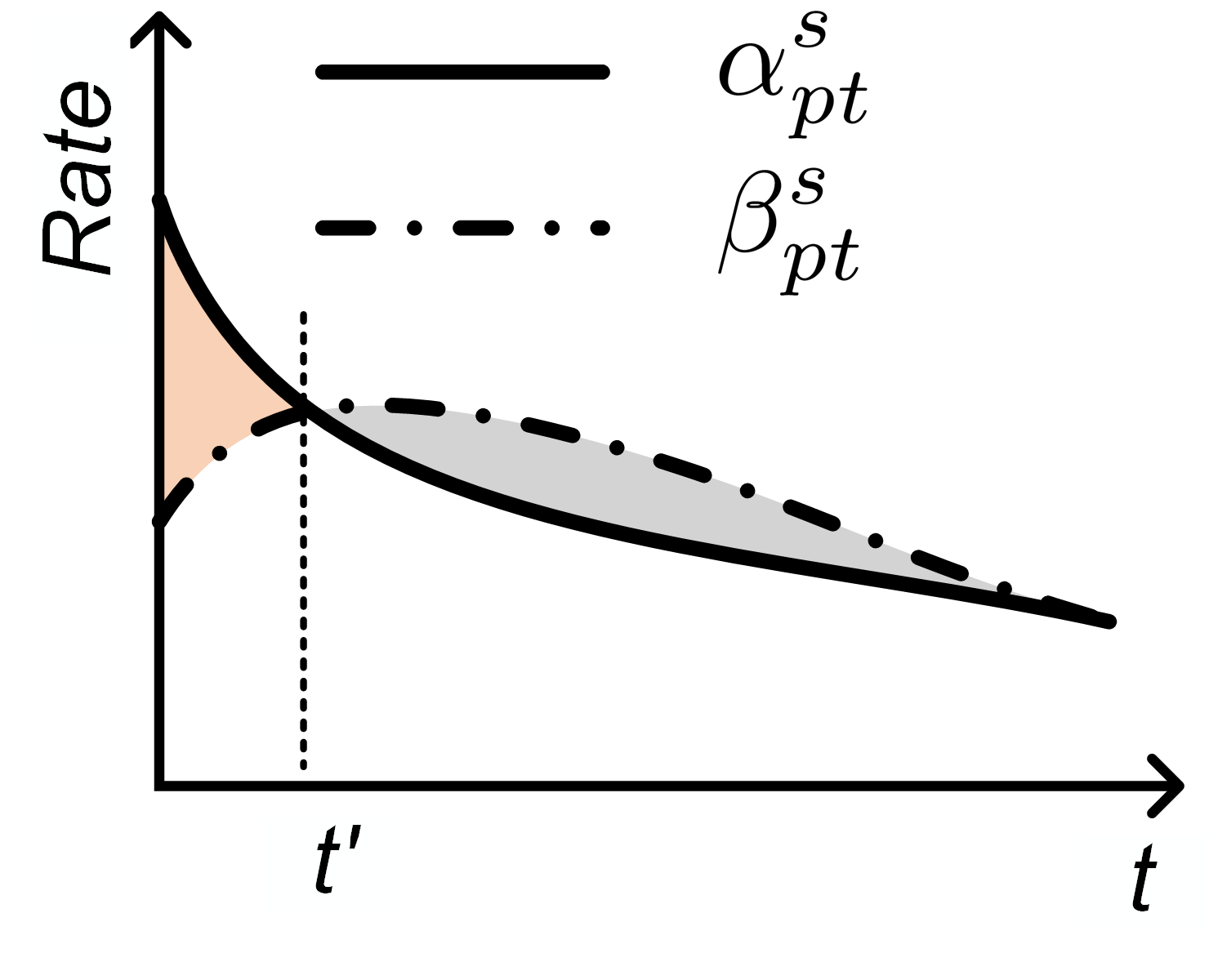}}
\caption{Time-varying process of the arrival and departure rates. (a) $\alpha_{pt}^s<\beta_{pt}^s$ at the beginning. (b) $\alpha_{pt}^s>\beta_{pt}^s$ at the beginning.}
\label{fig3}
\end{figure}

When $\frac{n}{\mu}\sum_{p\in \mathcal{P}}\sum_{s=1}^{|p|}\sum_{k=1}^s\Phi_{pe}^sR_{pt}^k$ is positive, the third term of price update in (\ref{Corollary1}) is usually negligible. Then, $\lambda_{et}$ increases, making related $x_{jp}^*$ and $\alpha_{pt}^1=\sum_{j\in\mathcal{J}_t}x_{jp}^*\Theta_{jp}$ decrease. Based on the aforementioned deduction, the values of $\beta_{pt}^1$ and $R_{pt}^1$ decrease after an inflection time $t^1$. Since $\alpha_{pt}^2=\beta_{pt}^1$ decreases after $t^1$, the values of $\beta_{pt}^2$ and $R_{pt}^2$ decrease after an inflection time $t^2$. This process continues until all $R_{pt}^s,s=1...|p|$ decreases after $t^{|p|}$. As all $R_{pt}^s$ decrease to 0, $\lim_{t\rightarrow \infty}\hat{\lambda}_{et}=\lambda^*_{et}$.  Then, the derivative of $\hat{\lambda}_{et}$ satisfies $\lim_{t\rightarrow \infty}\frac{d\hat{\lambda}_{et}}{dt}=\frac{d\lambda^*_{et}}{dt}\approx 0$ since the fluctuation of $\lambda^*_{et}$ is much slower than the iteration convergence process. In this situation, the second term in (\ref{Ft}) is negligible.

As a result, we get $\Delta \mathcal{F}_t\le 0$ and $\lim_{t\rightarrow \infty} \mathcal{F}_t = 0$, which means $\lim_{t\rightarrow \infty}\lambda_{et}=\hat{\lambda}_{et}$. Finally, we get
\begin{equation}
\lim_{t\rightarrow \infty}\lambda_{et}=\hat{\lambda}_{et}=\lambda_{et}^*,\quad \forall e.
\end{equation}
\end{proof}

Theorem 1 reveals the convergence of the iteration strategy (\ref{Corollary1}) under the condition that the optimal prices $\lambda^*_{et}$ fluctuate slowly. However, this condition may not always hold for real networks when the traffic matrix is rapidly time-varying. In such circumstances, the iteration strategy cannot guarantee a strict convergent state, where $R_{pt}^s=0,\forall p,s$ and $\lambda_{et}=\lambda^*_{et},\forall e$. Instead, the network reaches a dynamic balance, where each $\lambda_{et}$ lags behind the corresponding $\lambda^*_{et}$ and each $R_{pt}^s$ fluctuates within a specific range. We provide numerical results in Sec. VI-B to show this dynamic balance. The flow queueing delay is nonzero if $R_{pt}^s\neq 0$, and long flow queueing delay may result in unsatisfactory E2E delay for delay sensitive flows and TCP timeout. To constrain the flow queueing delay in a specific range, we utilize a dynamic step size mechanism (see Sec. V).
\section{System Design and Algorithms}
Based on the theorems proposed in Sec. IV, we propose a system design of the asynchronous TE paradigm. 

As introduced in Sec. III-B, a path selector is employed by each service broker for routing. The path selector utilizes two routing policies for delay-sensitive and delay-tolerant flows, as in Algorithm 1. For delay-sensitive flows, an increase in E2E delivery delay may cause significant user experience degradation. The E2E delivery delay is the sum of link propagation delays and queueing delays. The link propagation delay $d_{et}$ depends on the link length, usually a fixed value. In addition, the queueing delay $q_{et}$ of link $e$ can be estimated using Little's Law as
\begin{equation}
q_{et}=\frac{\sum_{p\in\mathcal{P}}\sum_{s=1}^{|p|}\Phi_{pe}^sR_{pt}^s}{\mathcal{C}_e}.
\label{queueingdelay}
\end{equation}
Then, a shortest-delay path strategy is used based on $d_{et}$ and $q_{et}$ as in line 6 of Algorithm 1. For delay-tolerant flows, users are more concerned about throughput than the E2E delivery delay. The maximum throughput is achieved when the network allocates bandwidth on the lowest-price path of a flow. Consequently, a lowest-price path strategy is utilized, as shown in line 8 of Algorithm 1. After selecting a path, the service broker calculates the bandwidth allocation $x_{jp}^*$ and configures $\langle j, \Theta_{jp}, x_{jp}^*\rangle$ to the meter of corresponding edge node. The meter is responsible for tracking the traffic rates of flows and discarding the packets that exceed the allocated bandwidth. The complete FDTC action can be depicted using the pseudo-code in Algorithm 1. 
\begin{algorithm}[htb]
\caption{FDTC Action}
\hspace*{0.02in} {\bf Input: } The packet of a new flow, $\mathcal{P}$, $\Phi$, $d_{et},q_{et},\forall e\in \mathcal{E}$\\
\hspace*{0.02in} {\bf Output:} $\Theta_{jp}$, $x_{jp}^*$
\begin{algorithmic}[1]
\STATE The packet classifier executes flow table lookup to find the identifier $j$ for the packet.
\STATE Export the utility function $u_j(\cdot)$ and QoS demand of $j$ from the flow information database.
\STATE $\Theta_{jp}\leftarrow 0,\quad\forall p\in \mathcal{P}$
\STATE $x_{jp}\leftarrow 0,\quad\forall p\in \mathcal{P}$
\IF {flow $j$ is delay sensitive}
  \STATE $p=\mathop{\arg\min_{p\in \mathcal{P}}}\sum_{e\in\mathcal{E}}\Phi_{pe}(d_{et}+q_{et})$
\ELSE
  \STATE $p=\mathop{\arg\min_{p\in \mathcal{P}}}\sum_{e\in\mathcal{E}}\Phi_{pe}\lambda_{et}$
\ENDIF
\STATE $\Theta_{jp}=1$
\STATE $x_{jp}^*=\mathop{\arg\max}_{x_{jp}\ge 0}\left(u_j(x_{jp})-\sum_{e\in \mathcal{E}}\Phi_{pe}\lambda_{et}x_{jp}\right)$
\STATE Configure a meter item $\langle j, \Theta_{jp}, x_{jp}^*\rangle$ for flow $j$. 
\RETURN $\Theta_{jp}$, $x_{jp}^*$.
\end{algorithmic}
\end{algorithm}

\begin{figure}[t]
\centering
\subfloat[]{\includegraphics[height=1.8in]{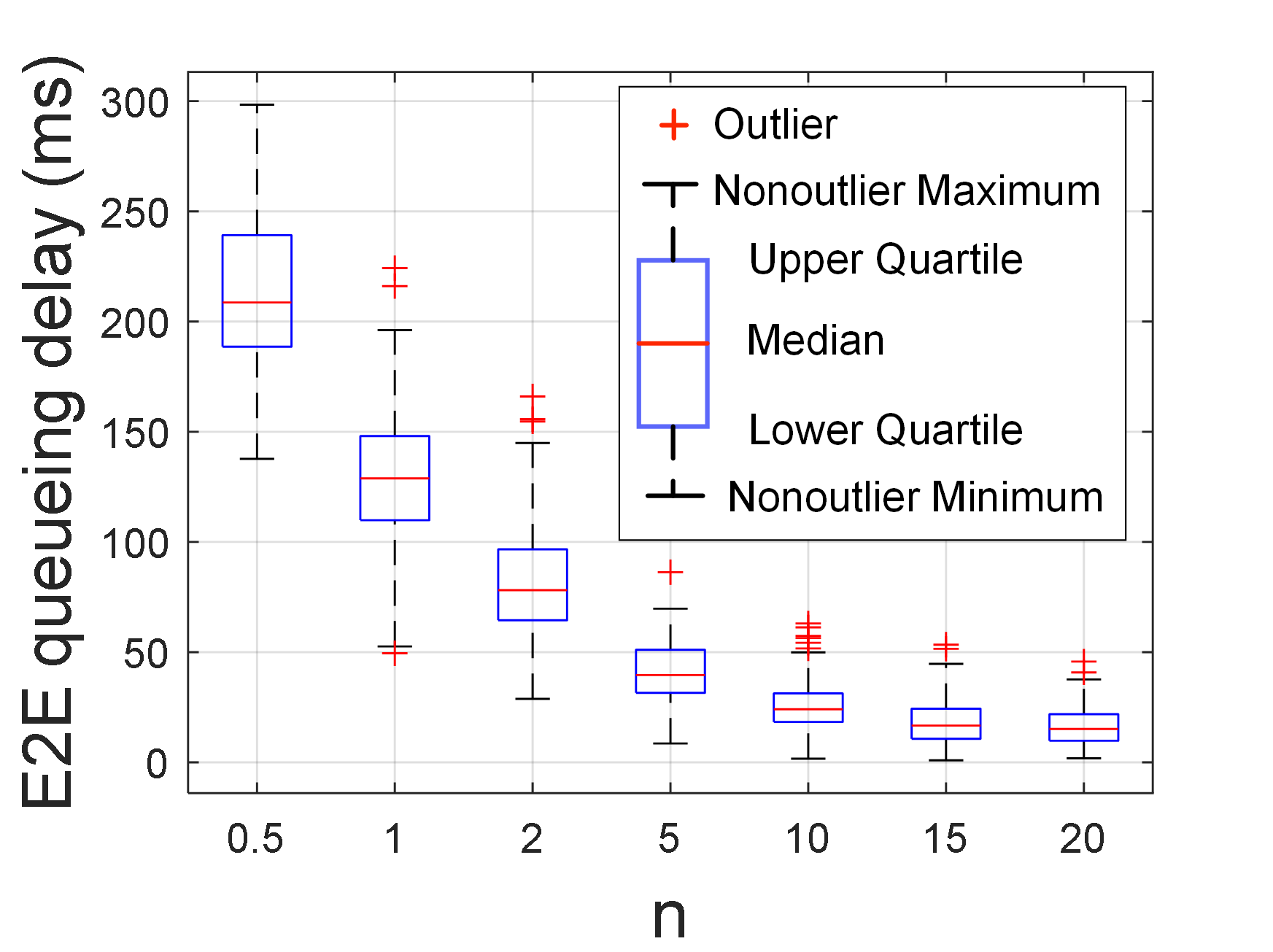}}\\
\subfloat[]{\includegraphics[height=1.8in]{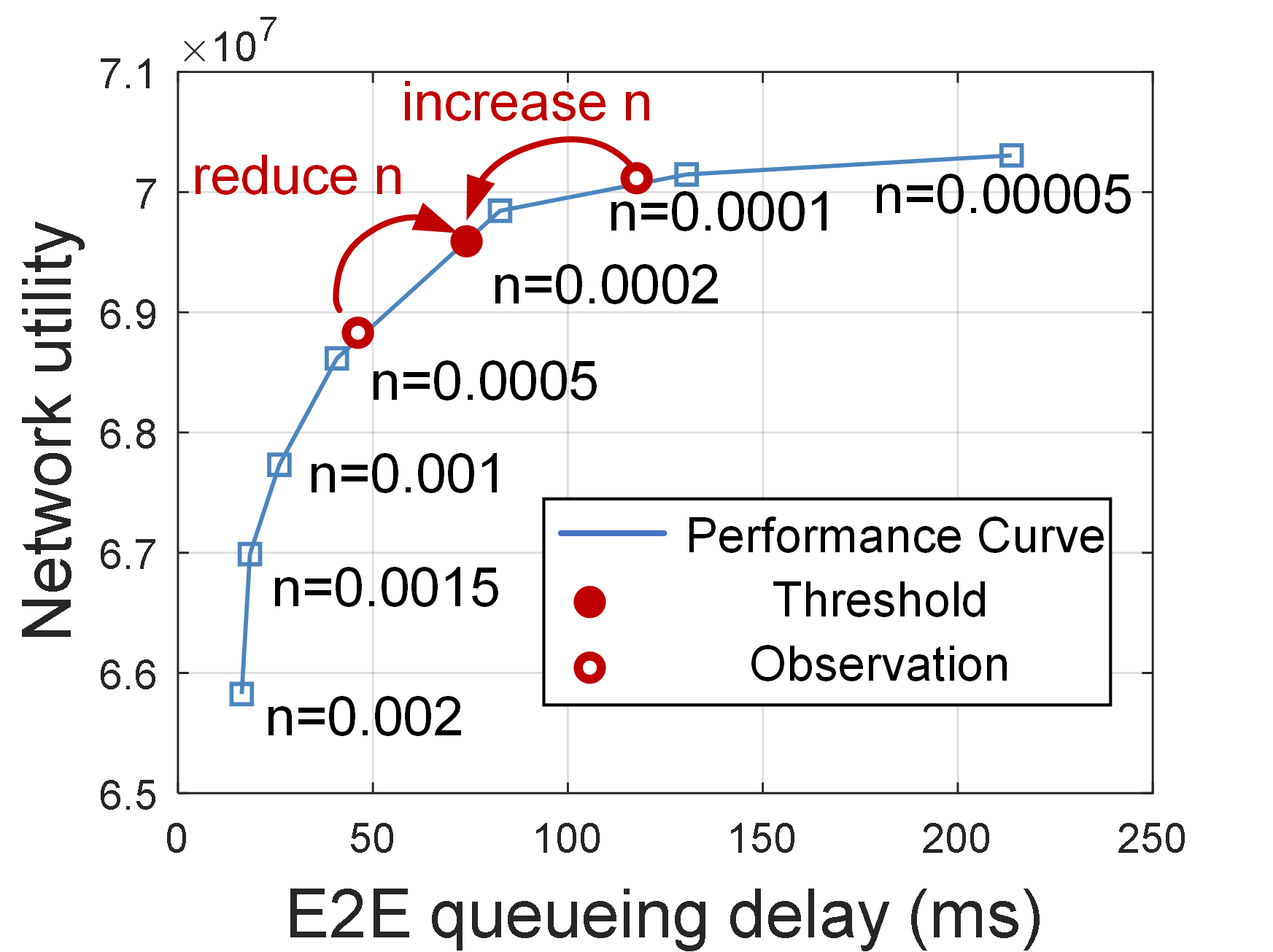}}
\caption{Impact of $n$ on E2E queueing delay and network utility. (a) Queueing delay distribution. (b) Performance curve under different values of $n$. More simulation details are given in Sec. VI.}
\label{fig5}
\end{figure}
The prices and delay parameters used in the FDTC action are provided and updated by the TE server. Initially, a traffic rate monitor is deployed in each core switch, which records the traffic rate $r_{pt}^s$, data volumes $R_{pt}^s$ in $Q_p^s$ of each associated path $p$, and idle bandwidth $\mathcal{I}_{et}$. The core switches upload $r_{pt}^s, R_{pt}^s, \mathcal{I}_{et}$ to the TE server when a preset condition of link price update is met. Afterwards, the TE server employs Algorithm 2 to calculate a new set of prices and configures them to underlying nodes with the assistance of the SDN controller.  
\begin{algorithm}[htb]
\caption{NIPU Action}
\hspace*{0.02in} {\bf Input: } $r_{pt}^s$, $R_{pt}^s$, $\mathcal{I}_{et}$, the price value $\lambda_{et}$ and step size $n_t$, $\Theta$, $\Phi$, the threshold $W^*$ of average E2E queueing delay, $\mathcal{C}_e$ and step size parameters $\varepsilon, \mu$\\
\hspace*{0.02in} {\bf Output:} $n_{(t+d)}$, $\lambda_{e(t+d)}$
\begin{algorithmic}[1]
\FOR{each $p\in\mathcal{P}$}
 \FOR{s=1 to $|p|$}
\STATE $d_{pt}^s= r_{pt}^{s\text{-1}}-r_{pt}^s \quad \#$  retention rate
\STATE $\mathcal{O}_{pt}^s=\sum_{k=1}^s d_{pt}^k\quad \#$ overload rate
\ENDFOR
\ENDFOR
\STATE $\mathbb{E}[q_{et}] = ({\sum_{p\in\mathcal{P}}\sum_{s=1}^{|p|}\Phi_{pe}^s\mathbb{E}[R_{pt}^s}])\mathcal{C}_e^{-1}$
\STATE $W_p = \sum_{s=1}^{|p|}\sum_{e\in\mathcal{E}}\Phi_{pe}^s \mathbb{E}[q_{et}]$
\STATE $\overline{W}=\frac{1}{|\mathcal{P}|}\sum_{p\in \mathcal{P}}W_p\quad\#$ average E2E queueing delay
\IF {$\overline{W}>W^*$}
\STATE $n_{t+d}= n_t + \varepsilon$
\ELSE
\STATE $n_{t+d}= n_t - \varepsilon$
\ENDIF
\FOR {each link $e\in \mathcal{E}$}
\STATE $\dot{\lambda}_{et}^{(1)} = \sum_{p\in\mathcal{P}}\sum_{s=1}^{|p|}\sum_{k=1}^{s}\Phi_{pe}^s R_{pt}^k$
\STATE $\dot{\lambda}_{et}^{(2)} = \sum_{p\in\mathcal{P}}\sum_{s=1}^{|p|}\Phi_{pe}^s\mathcal{O}_{pt}^{s}-\mathcal{I}_{et}$
\STATE $\lambda_{e(t+d)} \leftarrow \left[\lambda_{et} + n\dot{\lambda}_{et}^{(1)} + \mu \dot{\lambda}_{et}^{(2)}\right]^+$
\ENDFOR
\RETURN $n_{(t+d)}$, $\lambda_{e(t+d)}$
\end{algorithmic}
\end{algorithm}

Theorem 2 provides an iteration direction in (\ref{Corollary1}). In addition, the step size $n$, which reflects the algorithm's sensitivity to lengths $R_{pt}^s$ of virtual queues, has an impact on network performance. As we use a larger $n$, a link price $\lambda_{et}$ rises more rapidly when the term $\dot{\lambda}_{et}^{(1)}$ is positive, resulting in a faster drop in the arrival rates $\alpha_{pt}^s$ and $R_{pt}^s$ (see the proof of Theorem 2). This can be expressed as a positive relation between the drop speed $-\frac{\text{d}R_{pt}^s}{\text{d}t}$ and $n$. In this context, the average value $\mathbb{E}[R_{pt}^s]$ of $R_{pt}^s$ is negatively correlated with $n$ since 
\begin{equation}
\mathbb{E}[R_{pt}^s] = \int_0^\infty\frac{1}{t}\int_{0}^t \left(\frac{\text{d}R_{p\tau}^s}{\text{d}\tau}\right) \text{d}\tau\text{d}t.
\end{equation}
In addition, the average queueing delay $\mathbb{E}[q_{et}]$ of each link $e$ is positively correlated with $\mathbb{E}[R_{pt}^s]$ according to (\ref{queueingdelay}).
Then, the average E2E queueing delay $W_p$ along path $p$ is positively correlated with $\mathbb{E}[R_{pt}^s]$ since
\begin{equation}
W_p = \sum_{s=1}^{|p|}\sum_{e\in\mathcal{E}}\Phi_{pe}^s \mathbb{E}[q_{et}].
\end{equation}
As a result, the average E2E queueing delay is negatively correlated with $n$.
Fig. \ref{fig5}(a) shows the distribution of the average E2E queueing delay experienced by flows during a simulated experiment. The result verifies that the E2E queueing delay has a negative shift as $n$ increases. Meanwhile, the network encounters utility degradation as $n$ grows. Fig. \ref{fig5}(b) depicts the performance curve when different $n$ values are utilized during iteration. Both the E2E queueing delay and network utility decrease as $n$ increases. In real networks, we usually need to constrain the E2E queueing delay under a specific threshold $W^*$. To realize it, we find the optimal step size $n^*$ that makes the E2E queueing delay equal $W^*$. When a queueing delay shorter than the threshold is observed, the network reduces $n$ to enhance network utility (see line 12 of Algorithm 2). Conversely, the network raises $n$ to reduce the E2E queueing delay (see line 10 of Algorithm 2).
\section{Numerical Result}\label{NumericalResult}
\subsection{Simulation Setup}
\begin{figure}[h]
\centering
\includegraphics[width=2.8in]{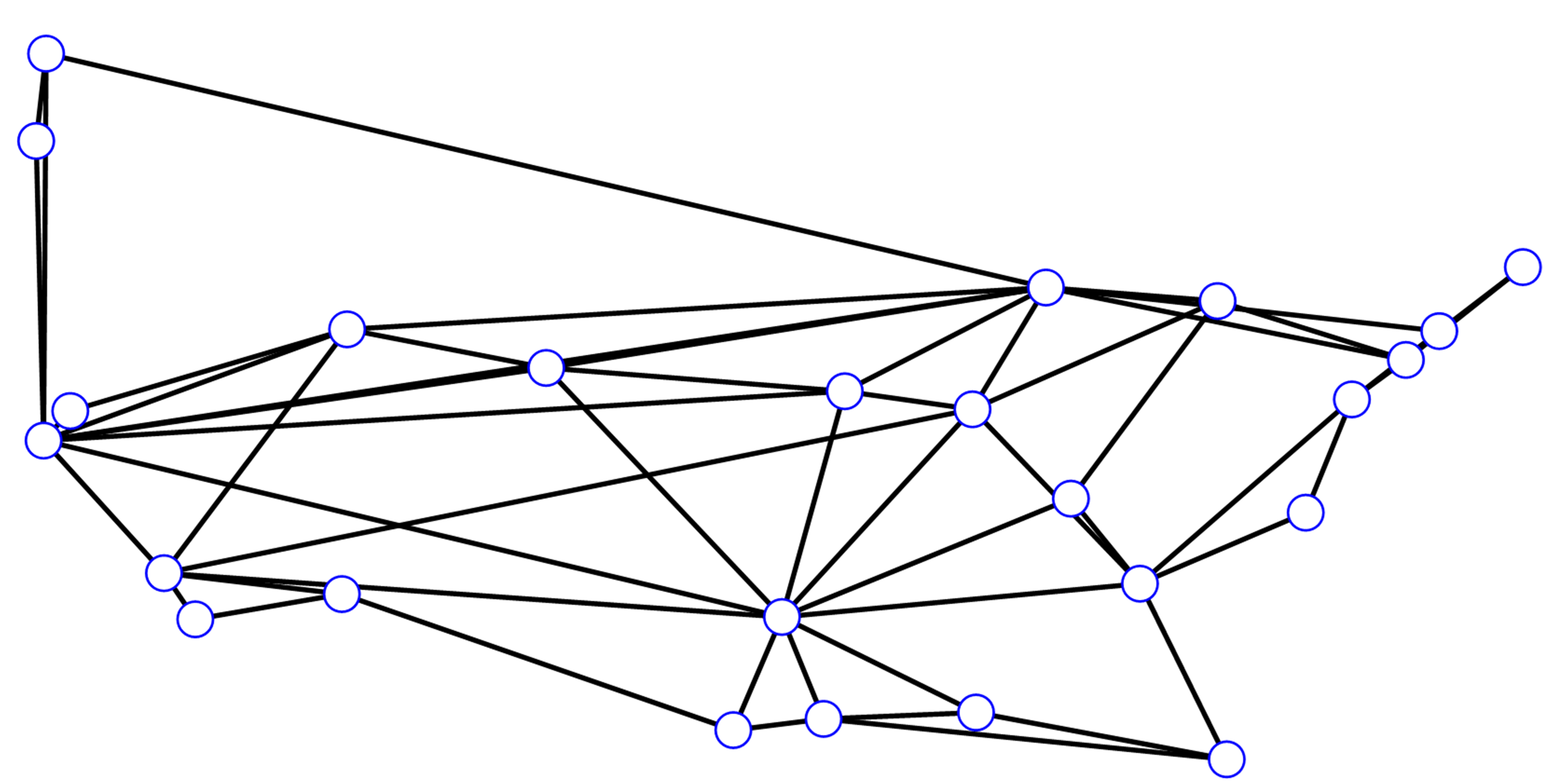}
\caption{Topology used in the simulation.}
\label{fig11}
\end{figure}

\textbf{Network Settings.} This section investigates the performance of AMTM and compares it with other existing approaches. We build a flow-level simulator based on Python, and Appendix B provides a detailed description of the simulator. The topology of a real network from an open source dataset\cite{GlobalNetRu} is utilized during simulation. Fig. \ref{fig11} illustrates this topology, consisting of 25 nodes and 110 directed links, whereby the link capacity is set as 5 Gbps. A service broker is deployed in each node, and a TE server is deployed in the network. The service brokers execute Algorithm 1 upon the arrival of new flows. The switch nodes generate observations $r_{pt}^s$, $R_{pt}^s$, and $\mathcal{I}_{et}$ and send them to the TE server. The TE server is set to execute Algorithm 2 once in a second.

\textbf{Traffic Settings.} During the simulation, users generate multi-class flows based on the parameters specified in Table \ref{Flow}. Interactive flows, which are short and delay-sensitive, have the highest weight, since little QoS degradation causes a poor user experience. Similarly, streaming media flows with higher rates and durations require a low delivery delay to enhance the user experience. Elastic flows, such as FTP data transfers, are delay-tolerant and have the lowest weights. We maintain the ratio of delay-sensitive flows in both quantity and traffic, in compliance with the traffic characteristics in previous research \cite{benson2010network}.
To guarantee a fair service among different flows, we utilize $u_j(x_j)=\frac{w_j}{1-\alpha}x_j^{1-\alpha}$ as the utility function, and the corresponding $w_j$ are listed in Table \ref{Flow}. The value of $\alpha$ indicates the degree of marginal diminishing in throughput, and we set $\alpha=0.5$, a value located between $\alpha=0$ (linear utility function) and $\alpha=1$ (log function). The arrival of the generated flows follows a Poisson process, and the arrival intensity is adjusted to various values to investigate the performance under different traffic loads. In addition, each generated flow is randomly assigned a source node and a destination node. 
\begin{table}[h]
\caption{Parameters of multiclass flows.}
\label{Flow}
\centering
\begin{tabular}{p{1.6cm}<{\centering} p{1.8cm}<{\centering} p{1.7cm}<{\centering} p{1.7cm}<{\centering}}
\toprule
 & Interactive\newline Flow & Streaming\newline Media Flow & Elastic \newline Flow \\ \midrule
Bandwidth & 10 Mbps & 20 Mbps & 100 Mbps \\ \hline
QoS Demand & Low delay & Low delay & Throughput\\\hline
Duration Range & 10 to 30\newline seconds & 1 to 5\newline minutes & 10 seconds \newline to 10 minutes \\\hline
Weight $w_j$ & 3 & 2 & 1\\\hline
Generation Probability & 86$\%$ & 7$\%$ & 7$\%$\\\hline
Traffic Ratio &7$\%$& 10$\%$ & 83$\%$ \\\bottomrule
\end{tabular}
\end{table}

\textbf{Queuing Mechanism Settings.} To guarantee a low delivery delay for interactive and streaming media flows, switch nodes employ a two-priority queuing mechanism for each egress link. The packets of interactive and streaming media flows are buffered in a high-priority queue and are always sent first. The packets of elastic flows, on the other hand, are buffered in a low-priority queue, and their packets are sent only when the high-priority queue is empty. 
\subsection{Convergence of AMTM}
\begin{figure}[t]
\centering
\subfloat[]{\includegraphics[height=1.35in]{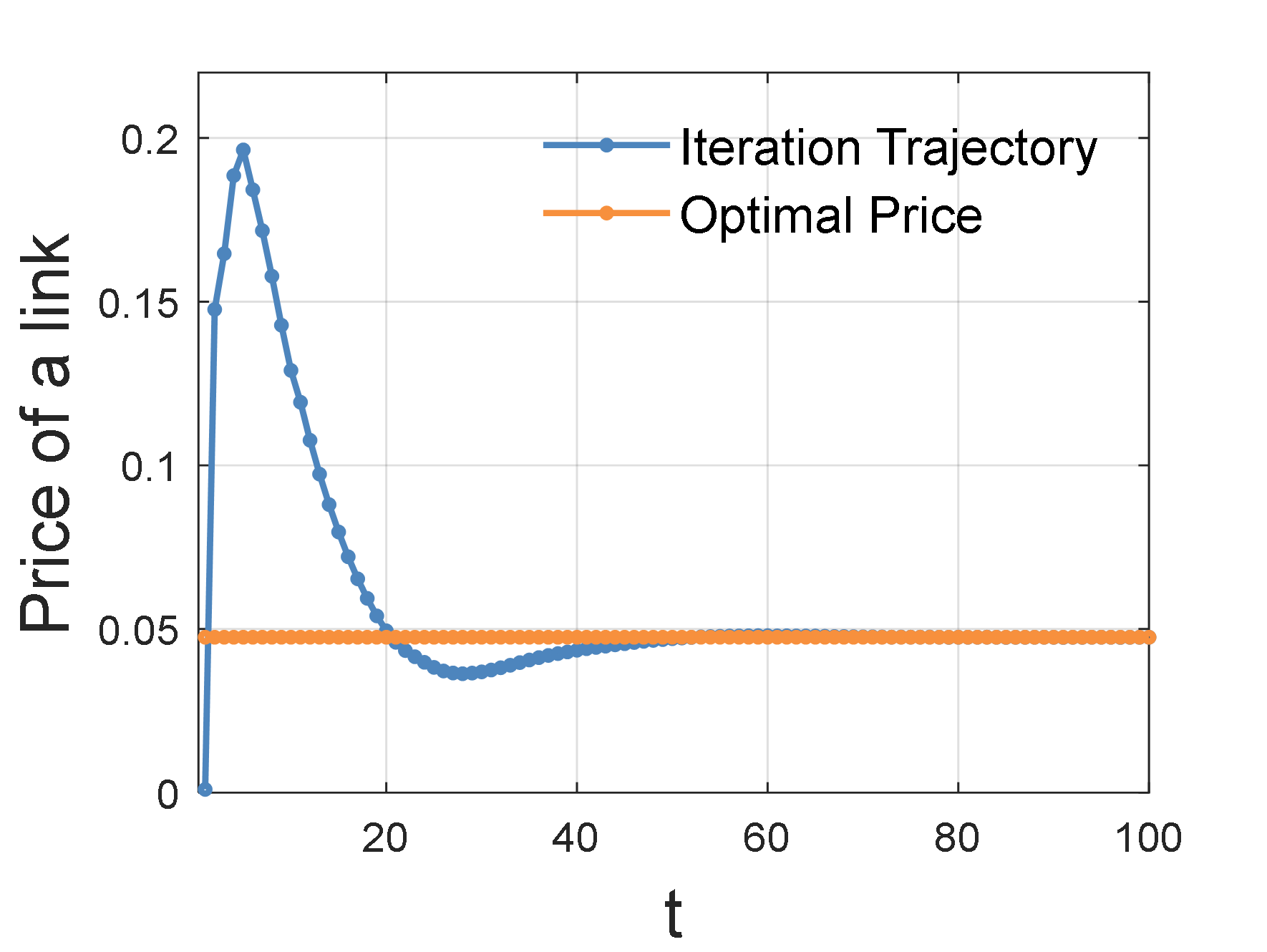}}
\subfloat[]{\includegraphics[height=1.35in]{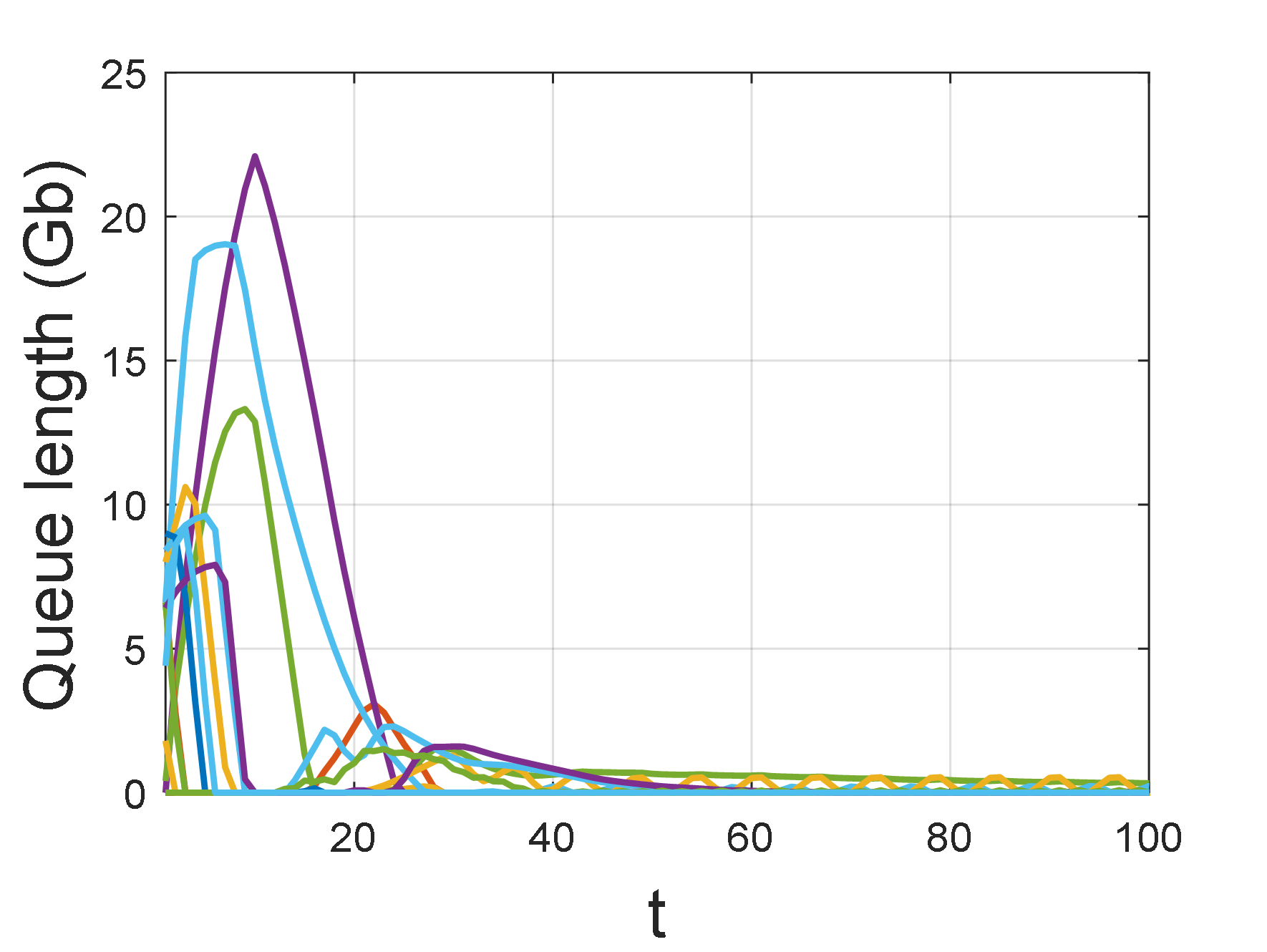}}\\
\subfloat[]{\includegraphics[height=1.35in]{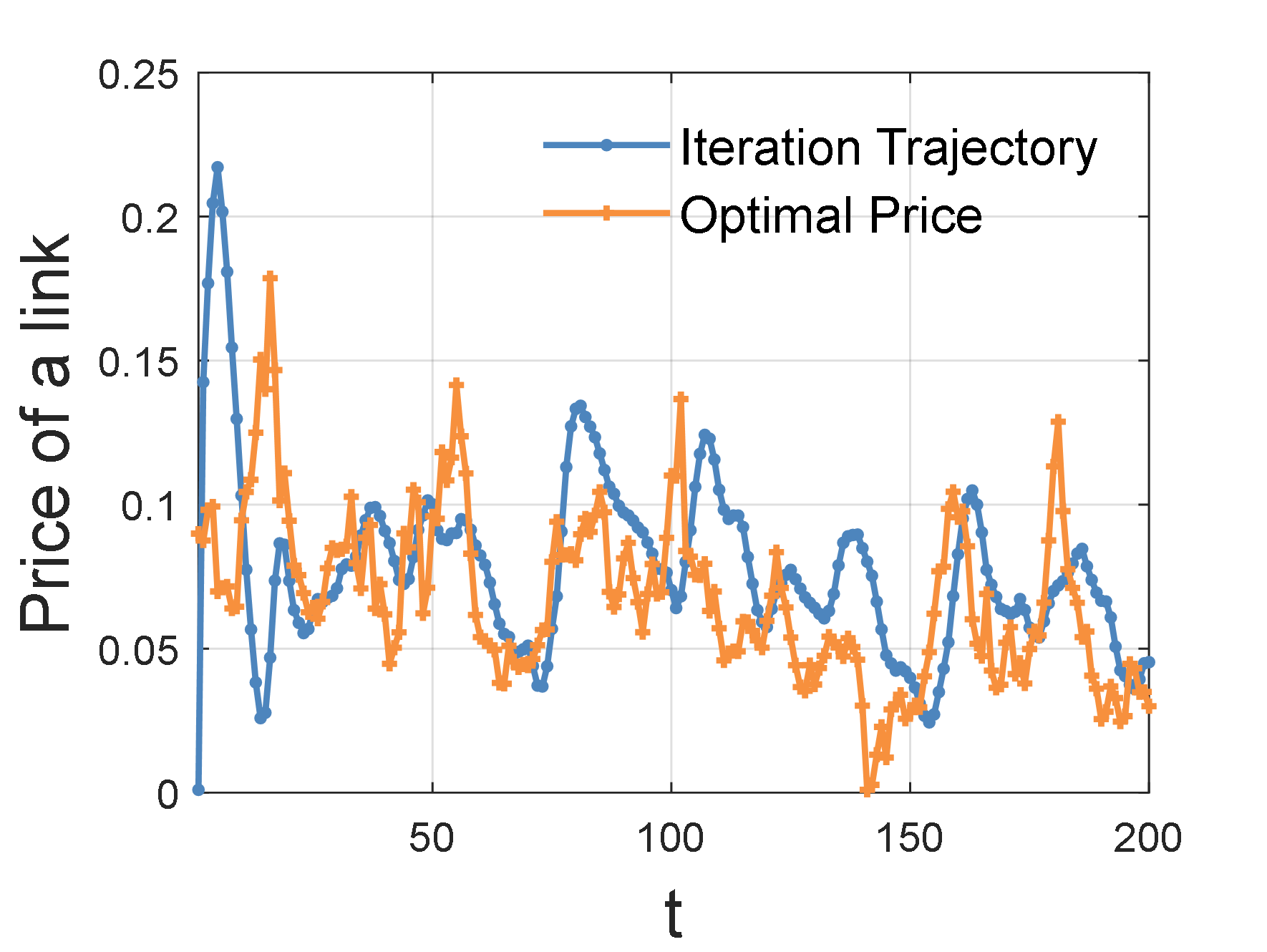}}
\subfloat[]{\includegraphics[height=1.35in]{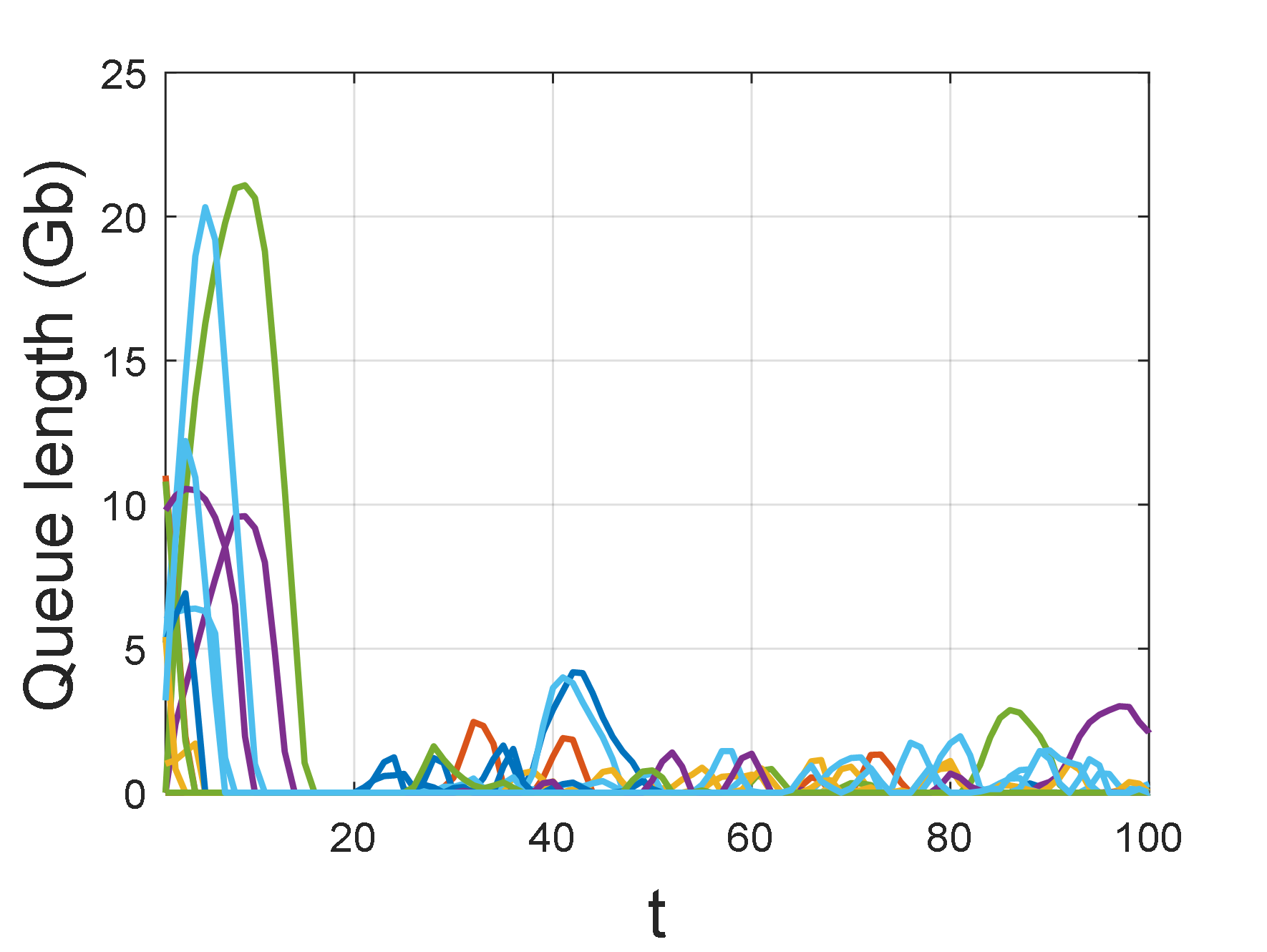}}
\caption{Convergence of AMTM. (a) Price of a link under a stationary traffic input. (b) Queue lengths under a stationary traffic input. (c) Price of a link under a time-varying traffic input. (d) Queue lengths under time-varying traffic input.}
\label{fig4}
\end{figure}

In Sec. IV, we prove that the iteration strategies proposed in (\ref{Corollary1}) make link prices converge to the optimal prices when the fluctuation of the optimal prices is slow. The effectiveness of the strategies is confirmed through simulation, where we input 10,000 stationary flows with fixed parameters. Given the fixed value of flow parameters $\mathcal{J}_t$ and $\mathbf{u}_t$, the optimal prices $\lambda^*_{et}$ of $\mathcal{U}(\mathcal{J}_t,\mathbf{u}_t,\Theta,\Phi,\mathcal{C})$ remain unchanged. As depicted in Fig. \ref{fig4}(a), AMTM makes the price iteration trajectory $\lambda_{et}$ of a link converge to $\lambda^*_{et}$. Meanwhile, we recorded the physical queue lengths of 20 randomly selected links, as depicted in Fig. \ref{fig4}(b). The result shows that the physical queue length of each link gradually converges to zero. 

However, actual networks usually experience nonstationary traffic inputs due to the continuous arrival of new flows, resulting in time-varying optimal prices $\lambda^*_{et}$. Therefore, the network cannot ensure that $\lambda^*_{et}$ fluctuates at a slower rate compared to the iteration convergence process. In this case, the condition stipulated in Theorem 2 cannot be satisfied. To investigate the network states under this situation, we randomly generate flows with a 30 s$^\text{-1}$ Poisson arrival intensity and input them into the simulator. Fig. \ref{fig4}(c) displays the optimal price $\lambda^*_{et}$ and iteration trajectory $\lambda_{et}$ of a link in this case, where it is evident that the iteration trajectory $\lambda_{et}$ lags behind $\lambda^*_{et}$. Consequently, an error always exists between $\lambda_{et}$ and $\lambda^*_{et}$. However, this error does not lead to diverging queue lengths or queueing delays. We observed the physical queue lengths of the same 20 links, and they fluctuated around zero, as illustrated in Fig. \ref{fig4}(d), which could be considered a "dynamic balance." As a result, the flows experience non-zero but finite E2E queueing delays.
\subsection{Performance Evaluation}
This section compares existing TE approaches and AMTM in terms of link utilization, network utility, delay, and scalability. We maintain a standardized setup during simulations to ensure a fair comparison. First, we set an interaction period of one second between the data plane of switching nodes and the TE server. Therefore, the TE period for both centralized and hierarchical TE approaches is one second. It cannot be decreased further due to the shortest algorithm running time of available algorithms (i.e., approximately one second\cite{xu2023teal}), control message delay, and rule configuration time. In certain instances, however, the TE period might be longer (e.g., several minutes), resulting in longer E2E delivery delays for flows. 
Additionally, we employ Yen's algorithm\cite{YenS} to find the top five shortest paths between each pair of nodes as the candidate paths. The compared approaches are detailed below.

\hangafter 1 
\hangindent 1em 
\noindent 
$\bullet$ \textbf{Centralized Scheme}. As introduced in Sections I and II, centralized TE follows a periodic paradigm and is widely used in data center networks\cite{10.1145/3230543.3230545,RN33,10.1145/2534169.2486012,7299623}. When these centralized TE approaches are utilized in general WANs, the traffic waits an average of 0.5 TE periods to acquire resources. Within each TE period, the TE server directly solves the network utility maximization problem in (\ref{ONU}) according to the demand of all arrived flows. Since link overloads are effectively eliminated, the average E2E delivery delay, excluding propagation delay, is 500 ms.

\hangafter 1 
\hangindent 1em 
\noindent 
$\bullet$ \textbf{Hierarchical Scheme Based on Resource Preallocation}. The TE server preallocates resources to each flow group consisting of flows between a specific pair of nodes based on the demand in the previous TE period. Subsequently, service brokers execute real-time routing and traffic control for these flows. This method is explored in references \cite{6678113,10.1145/2534169.2486012,9110786}. During the simulation, interactive and real-time multimedia flows are routed to the shortest path, while elastic flows are transmitted through multiple candidate paths, utilizing residual bandwidth. Upon arrival, some flows acquire receive resources immediately and begin transmission using the preallocated resources. However, preallocated resources do not always perfectly match traffic demands, so some flows have to wait for the next TE period.

\hangafter 1 
\hangindent 1em 
\noindent 
$\bullet$ \textbf{AMTM}. As presented in Sec. V, the average E2E queueing delay and the network utility achieved by AMTM are dependent on the step size $n$, which is adjusted based on a specific threshold of average E2E queueing delay. For this simulation, the threshold is set as 200 ms, and $\mu$ is set as 0.0001. The average E2E delays, excluding propagation delay, approximates the E2E queueing delay because the control message round-trip time and decision making time of FDTC is negligible.

A multitude of previous studies have demonstrated that distributed schemes, such as OSPF and ECMP, exhibit unsatisfactory performance when compared to the latest approaches mentioned above. Hence, performance evaluation of these distributed schemes has been omitted from this section. The simulation results with regard to various metrics are provided as follows. 

\begin{figure}[t]
\centering
\subfloat[]{\includegraphics[height=1.35in]{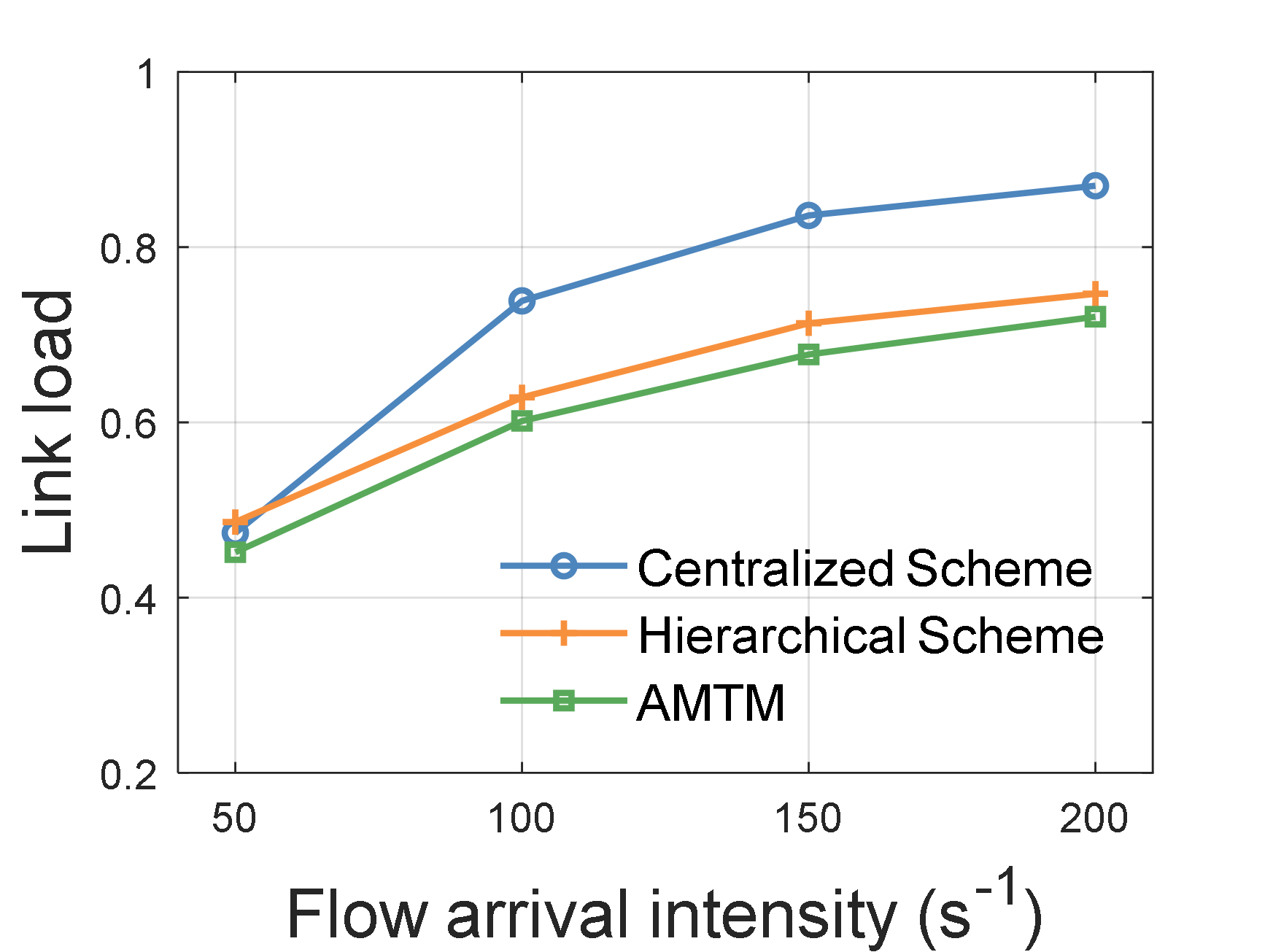}}
\subfloat[]{\includegraphics[height=1.35in]{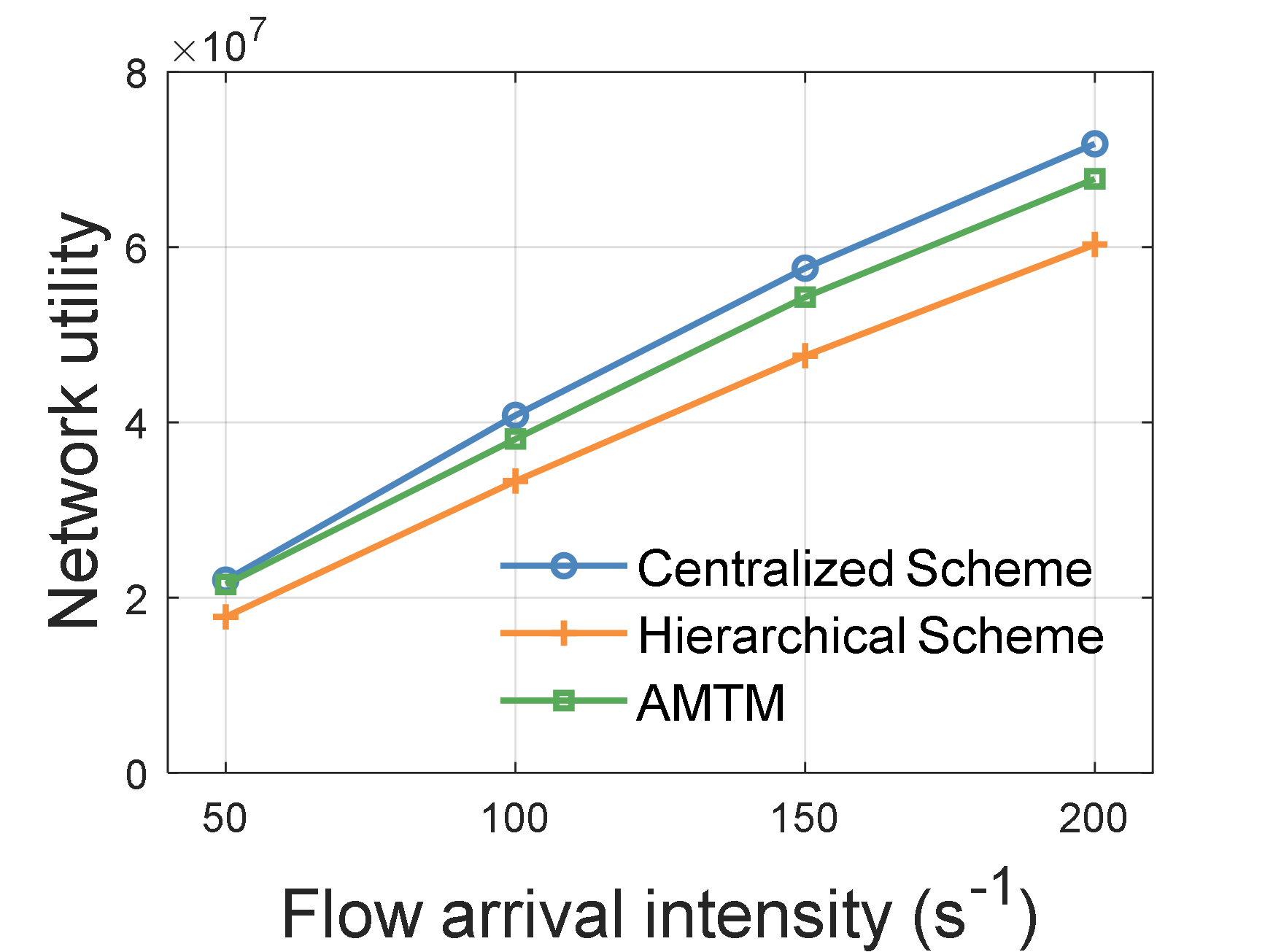}}\\
\subfloat[]{\includegraphics[height=1.35in]{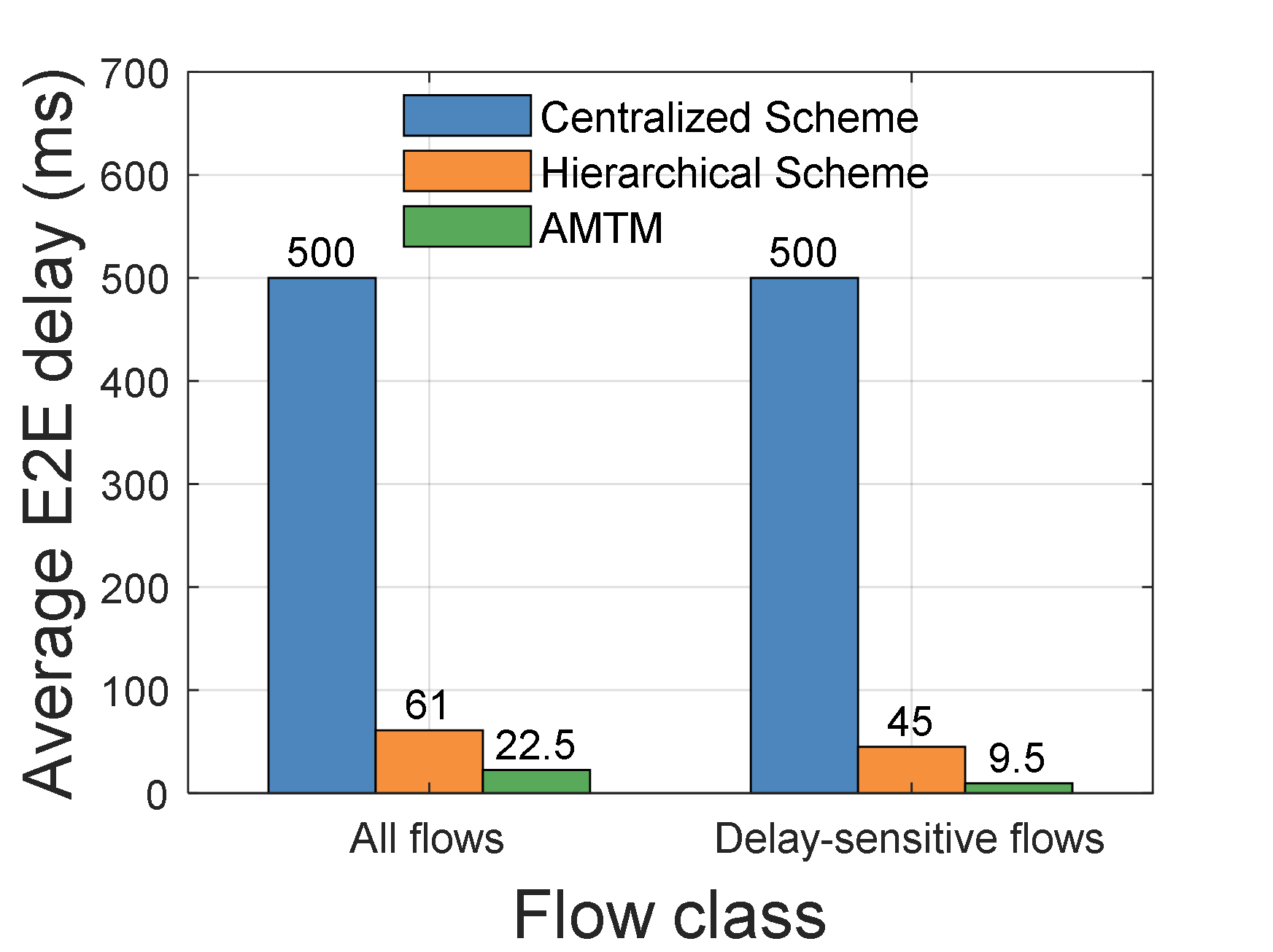}}
\subfloat[]{\includegraphics[height=1.35in]{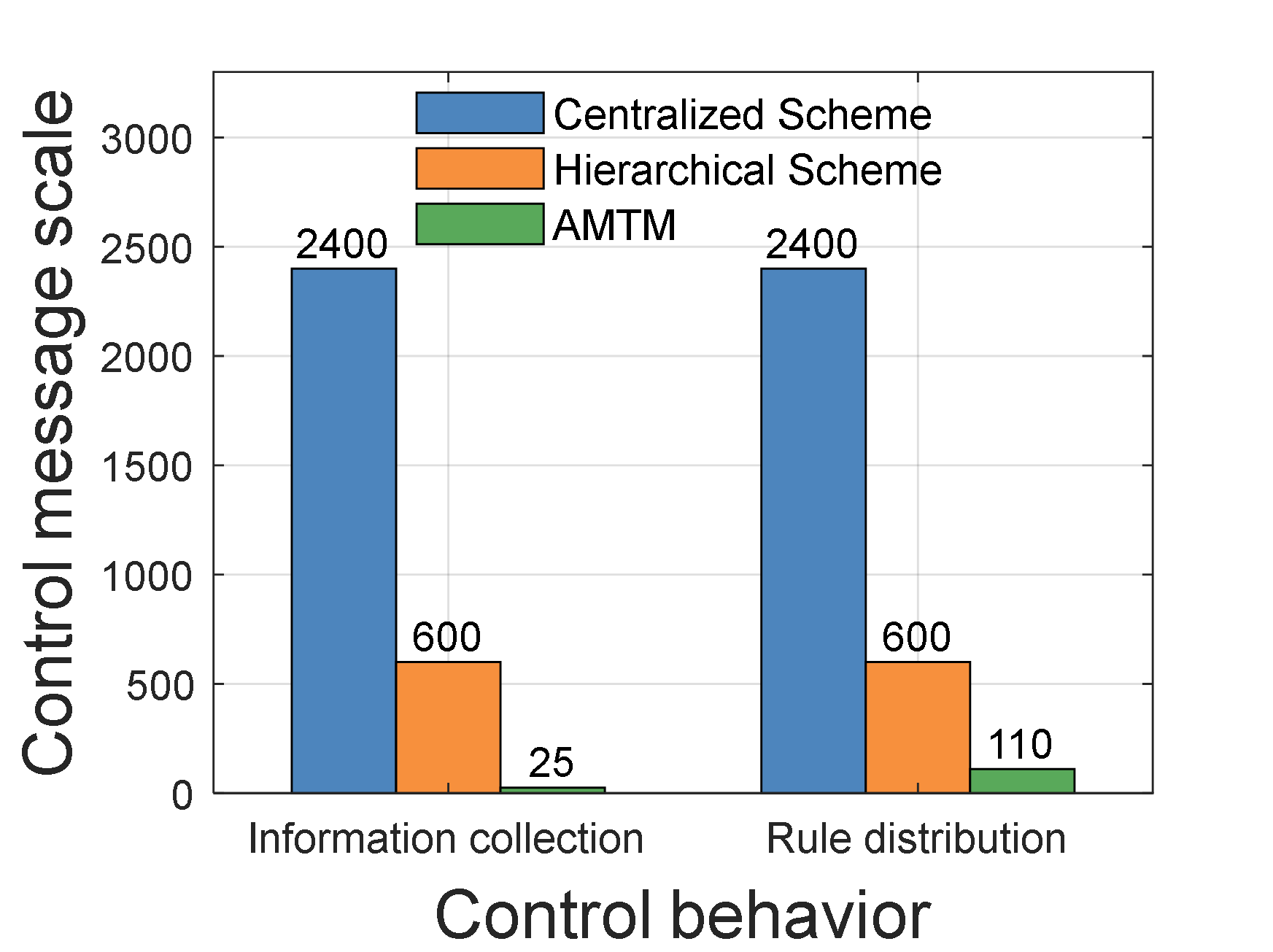}}
\caption{Simulation results. (a) Link load. (b) Network utility. (c) Average E2E delay, excluding propagation delay. (d) Control message scale.}
\label{fig12}
\end{figure}

\textbf{Link Load.} Link load, which is the average ratio of used bandwidth in all links, can provide insight into the degree of traffic load. As shown in Fig. \ref{fig12}(a), all three schemes exhibit an increase in link load as traffic arrival intensity increases. However, this increase slows down due to link saturation under heavy traffic loads. 

\textbf{Network Utility.} Fig. \ref{fig12}(b) displays the achieved network utility within 500 TE periods. AMTM achieves 12-20 $\%$ higher network utility than the hierarchical scheme. Furthermore, the network utility achieved by AMTM is near the maximum network utility achieved by the centralized scheme, with a gap of 2-7 $\%$.

\textbf{E2E Delay.} Fig. \ref{fig12}(c) displays the average E2E delays, excluding propagation delay\footnote{It should be noted that propagation delay varies widely depending on the geographical distance between nodes, ranging from tens of milliseconds to several seconds in different WANs. We exclude it from performance evaluation as it does not reflect the performance of a TE scheme.}, of all flows and delay-sensitive flows when the flow arrival intensity is 200 $\text{s}^\text{-1}$. The result indicates that AMTM and the hierarchical scheme can significantly reduce the average E2E delay to several tens of milliseconds. AMTM achieves the lowest delay among the three schemes, and the delay of delay-sensitive flows is less than 10 ms.  

\textbf{Scalability.} To compare the scalability of different schemes, we measure the number of control messages generated during the information collection and rule configuration phases in each period. Control message scale can provide an estimate of the control overhead, even though message formats may differ across systems. The centralized scheme gathers flow information and enforces flow-level rules, resulting in a control message scale of $o(\# \text{flow})$. The hierarchical scheme gathers the information of flow groups and deploys rules for them, resulting in a control message scale of $o(\# \text{flow group})$. In contrast, AMTM solely collects node states and distributes link prices, resulting in control message scales of $o(\# \text{node})$ and $o(\# \text{link})$, respectively. As illustrated in Fig. \ref{fig12}(d), AMTM exhibits the least control overhead in the simulated network. More importantly, the control overhead of AMTM does not increase as the number of flows grows, thus enhancing its scalability in large WANs.

In summary, AMTM reduces both the average queueing delay and control overhead compared to the existing schemes. Additionally, it outperforms the hierarchical scheme with a 12-20$\%$ improvement in network utility, while maintaining a close performance to the maximum network utility.
\section{Conclusion}
This paper presents an asynchronous multi-class traffic management scheme, known as AMTM. An asynchronous TE paradigm is established whereby service brokers execute local traffic control with a short delay at the network edge, and the TE server updates link prices to eliminate the decision conflicts between service brokers.
Additionally, a pricing strategy based on virtual queues in intermediate nodes is proposed for the long control loop in the asynchronous TE paradigm. Furthermore, this paper presents a system design and AMTM algorithms that utilize a dynamic step size mechanism. The simulation results demonstrate that the AMTM algorithms lead to convergence and effectively reduce E2E delay.

In future studies, it may be beneficial to utilize a specific link price set and iteration strategy for each flow class, as flows possess distinct features. Furthermore, exploring the use of the asynchronous paradigm in inter-network scenarios is a promising direction, as synchronous actions are usually impractical in multiple autonomous networks.


\appendices
\section{The Pricing Strategy Based on Dual-Decomposition}
This section introduces a price update strategy based on dual theory. According to dual theory, the Lagrange dual problem of $\mathcal{U}(\mathcal{J}_t,\mathbf{u}_t,\Theta,\Phi,\mathcal{C})$ is 
\begin{equation}
\begin{aligned}
&\text{minimize: }\varphi(\lambda), \text{ subject to }{\lambda_{e}\ge 0}, \varphi(\lambda) =\\
& \max_{x_{jp}\ge 0}\sum_{j\in \mathcal{J}_t}u_j(\sum_{p\in\mathcal{P}}x_{jp})-\sum_{e\in \mathcal{E}}\lambda_{e}(\sum_{j\in \mathcal{J}_t}\sum_{p\in\mathcal{P}}x_{jp}\Theta_{jp}\Phi_{pe}-\mathcal{C}_e),\\
\end{aligned}
\label{DualProblem}
\end{equation}
where link prices $\lambda=\{\lambda_{e}| e\in \mathcal{E}\}$ are dual variables. Since the bandwidth values $\{x_{jp}|p\in\mathcal{P}\}$ of each flow are independent, we can transform the expression of $\varphi(\lambda)$ to
\begin{equation}
\begin{aligned}
\varphi(\lambda) &= \sum_{j\in \mathcal{J}_t}\varphi_j(\lambda)+\sum_{e\in \mathcal{E}}\lambda_e \mathcal{C}_e,\\
\varphi_j(\lambda)&=\max_{x_{jp}\ge 0}u_j(\sum_{p\in\mathcal{P}}x_{jp})-\sum_{e\in \mathcal{E}}\lambda_e\sum_{p\in\mathcal{P}}x_{jp}\Theta_{jp}\Phi_{pe}.\\
\end{aligned}
\end{equation}
The primal problem (\ref{ONU}) and the Lagrange dual problem (\ref{DualProblem}) have strong duality according to the Slater's condition (i.e., strong duality holds when the primal problem is convex and strictly feasible). Therefore, a feasible iteration strategy is as follows.
\\\textbf{Lemma A.1}: \emph{Problem $\mathcal{U}(\mathcal{J}_t,\mathbf{u}_t,\Theta,\Phi,\mathcal{C})$ converges to its optimal solution using the following iteration strategy}.
\begin{equation}
\begin{aligned}
&x_{jp}^* \leftarrow \mathop{\arg\max}_{x_{jp}\ge 0}u_j(\sum_{p\in\mathcal{P}}x_{jp})-\sum_{e\in \mathcal{E}}\lambda_e\sum_{p\in\mathcal{P}}x_{jp}\Theta_{jp}\Phi_{pe},\\
&\lambda_e \leftarrow \left[\lambda_e + \mu (\sum_{j\in \mathcal{J}_t}\sum_{p\in\mathcal{P}}x_{jp}^*\Theta_{jp}\Phi_{pe}-\mathcal{C}_e)\right]^+,
\end{aligned}
\label{LemmaA1}
\end{equation}
\emph{where parameter $\mu$ controls the step size during iteration, and $[\cdot]^+$ denotes the projection onto the nonnegative orthant}
\begin{proof}
To prove the convergence of the iteration strategy, we construct the following function of the iteration trajectory $\lambda_{ei}$ and optimal solution $\lambda_e^*$ of dual problem (\ref{DualProblem})
\begin{equation}
\mathcal{F}_i=\frac{1}{2}\sum_{e\in \mathcal{E}}(\lambda_{ei}-\lambda_e^*)^2,
\end{equation}
where $i$ represents the $i^\text{th}$ iteration. 

When the value in $[\cdot]^+$ is positive, the increment of price after iteration is $\Delta\lambda_{ei} = \mu (\sum_{j\in \mathcal{J}_t}\sum_{p\in\mathcal{P}}x_{jp}^*\Theta_{jp}\Phi_{pe}-\mathcal{C}_e)$ according to (\ref{LemmaA1}). Then, the increment of $\mathcal{F}_i$ after the $i^\text{th}$ iteration is 
\begin{equation}
\begin{aligned}
\Delta\mathcal{F}_i=\mu \sum_{e\in \mathcal{E}}(\lambda_{ei}-\lambda_e^*)(\sum_{j\in \mathcal{J}_t}\sum_{p\in\mathcal{P}}x_{jp}^*\Theta_{jp}\Phi_{pe}-\mathcal{C}_e).
\end{aligned}
\label{FF}
\end{equation}
According to (\ref{LemmaA1}), $x_{jp}^*$ is the solution to the maximum problem in the expression of $\varphi(\lambda)$ in (\ref{DualProblem}), which means
\begin{equation}
\begin{aligned}
&\frac{\partial \varphi(\lambda)}{\partial \lambda_{ei}}=(\mathcal{C}_e-\sum_{j\in \mathcal{J}_t}\sum_{p\in\mathcal{P}}x_{jp}^*\Theta_{jp}\Phi_{pe})+\sum_{j\in \mathcal{J}_t}\sum_{p\in \mathcal{P}}(\dot{u}_j(x_{jp}^*)\\
&-\lambda_e\Theta_{jp}\Phi_{pe})\frac{\partial x_{jp}^*}{\partial \lambda_e}=(\mathcal{C}_e-\sum_{j\in \mathcal{J}_t}\sum_{p\in\mathcal{P}}x_{jp}^*\Theta_{jp}\Phi_{pe}),\\
\end{aligned}
\label{varphi}
\end{equation}
where $\dot{u}_j(x)=\frac{\text{d}u_j(x)}{\text{d}x}$. According to (\ref{FF}) and (\ref{varphi}), we get
\begin{equation}
\begin{aligned}
\Delta\mathcal{F}_i &= - \mu \sum_{e\in \mathcal{E}}(\lambda_{ei}-\lambda_e^*)\frac{\partial \varphi(\lambda)}{\partial \lambda_{ei}}\\ 
&\le -\mu \sum_{e\in \mathcal{E}}[\varphi(\lambda)|_{\lambda_e=\lambda_{ei}}-\varphi(\lambda)|_{\lambda_e=\lambda_e^*}]\le 0,
\end{aligned}
\label{deltaF}
\end{equation}
where $\varphi(\lambda)|_{\lambda_e=\lambda_e^*}$ is the minimum value of $\varphi(\lambda)$ according to the definition of $\lambda_e^*$. The first inequality holds because $\varphi$ is a concave function\footnote{This conclusion can be proved in many ways. If you are interested in the proof, see https://math.stackexchange.com/questions/1374399/why-is-the-lagrange-dual-function-concave} of $\lambda_e$. 

When the value in $[\cdot]^+$ is negative, $0\ge\Delta\lambda_{ei} > \mu (\sum_{j\in \mathcal{J}_t}\sum_{p\in\mathcal{P}}x_{jp}^*\Theta_{jp}\Phi_{pe}-\mathcal{C}_e)=-\mu \frac{\partial \varphi(\lambda)}{\partial \lambda_{ei}}$. The concavity of $\varphi(\lambda)$ guarantees that $-\mu \frac{\partial \varphi(\lambda)}{\partial \lambda_{ei}}(\lambda_{ei}-\lambda_e^*)\le 0$, which means $\lambda_{ei}-\lambda_e^*\ge 0$ in this case. Then, we get $(\lambda_{ei}-\lambda_e^*)\Delta\lambda_{ei}\le 0$ and
\begin{equation}
\Delta\mathcal{F}_i = \mu \sum_{e\in \mathcal{E}}(\lambda_{ei}-\lambda_e^*)\Delta\lambda_{ei}\le 0.
\end{equation}

Thus, the iteration strategy makes $\mathcal{F}_i$ converge to its minimum value 0, which also means $\lambda_{ei}$ converges to $\lambda_e^*$.
\end{proof}
\section{Simulator}
The simulator is programmed in Python. It comprises five objects: Traffic Source, Virtual Queue, Physical Queue, Node, and Network. Each object has specific variables and functions defining its local parameters and behaviors. The table below presents the main behaviors of each object. 
\begin{table}[htb]
\caption{Behaviors of the objects}
\centering
\begin{tabular}{p{1cm}<{\centering} p{7cm}}
\toprule
\multicolumn{1}{c}{\textbf{Object}} & \multicolumn{1}{c}{\textbf{Behaviors}}\\ \midrule
Traffic Source & $\diamondsuit$ Generate a new flow with specific attributes, including bandwidth, weight, duration, and QoS demands\\
& $\diamondsuit$ Transmit a flow to the connected edge node\\
& $\diamondsuit$ Delete a completed flow\\ \hline
\multirow{2}{1cm}{\centering Virtual Queue} & $\diamondsuit$ Enqueue/Dequeue a specific volume of data\\
& $\diamondsuit$ Generate the observations of the intention rate and queue length \\ \hline
\multirow{4}{1cm}{\centering Physical Queue} & $\diamondsuit$ Create/Delete a virtual queue\\
& $\diamondsuit$ Enqueue (dequeue) data into (from) a virtual queue\\
& $\diamondsuit$ Drop overflow data and record the overflow rate\\
& $\diamondsuit$ Generate the observation of idle bandwidth on its egress link\\\hline
\multirow{6}{1cm}{\centering Node (With a service broker)} & $\diamondsuit$ Create/Delete a physical queue\\
& $\diamondsuit$ Enqueue an arrived flow into a physical queue\\
& $\diamondsuit$ Dequeue data from a physical queue and send it to a node\\
& $\diamondsuit$ Route a flow to a path and determine its rate\\
& $\diamondsuit$ Download the updated link prices and network topology\\
& $\diamondsuit$ Collect observations from physical queues and upload them\\ \hline
\multirow{4}{1cm}{\centering Network (With a TE server)} & $\diamondsuit$ Create a group of nodes and traffic sources \\ 
& $\diamondsuit$ Connect the nodes and traffic sources\\
& $\diamondsuit$ Update link prices based on the uploaded observations\\
& $\diamondsuit$ Simulate data transfer between nodes\\
\bottomrule
\end{tabular}
\end{table}

Realizing the last behavior of the Network object
is the simulator's most difficult aspect. When the arrival process and service time of a queueing network follow Poisson process and exponential distribution, respectively, Jackson Network theorems can be employed without simulation. However, in real networks with nonideal settings, simulating data transfer between nodes is necessary. This process can be modeled as a continuous queue state transition, using a differential equation:
\begin{equation}
\frac{\text{d} \mathbf{Q}(t)}{\text{d}t} = f(\mathbf{Q}(t))+\mathbf{A}(t),
\label{AppendxB1}
\end{equation}
where $\mathbf{Q}(t)$ is the state matrix of virtual queues. Each element in $\mathbf{Q}(t)$ represents the length of a virtual queue. The matrix $\mathbf{A}(t)$ represents the external input traffic from users, and $f(\cdot)$ is determined by the queueing mechanism and network topology. Unfortunately, $f(\cdot)$ is non-linear. Existing simulators choose varying levels of granularity to simulate this continuous process through discrete events. In packet-level simulators, the transfer of a packet is considered as a discrete event. Our simulator discretizes continuous time into small time slots of length $\tau$ and then uses these slots to discretize (\ref{AppendxB1}) as:
\begin{equation}
\mathbf{Q}(t+\tau)\leftarrow \mathbf{Q}(t) + (f(\mathbf{Q}(t)) + \mathbf{A}(t))\tau.
\end{equation}
When $\tau$ is sufficiently small, the simulation results converge to the actual results. To achieve this, we decreased the time slot during the simulation until the queue lengths reached convergence.

\ifCLASSOPTIONcaptionsoff
  \newpage
\fi
\bibliographystyle{IEEEtran}
\bibliography{reference.bib}

\vfill
\begin{IEEEbiography}[{\includegraphics[width=1in]{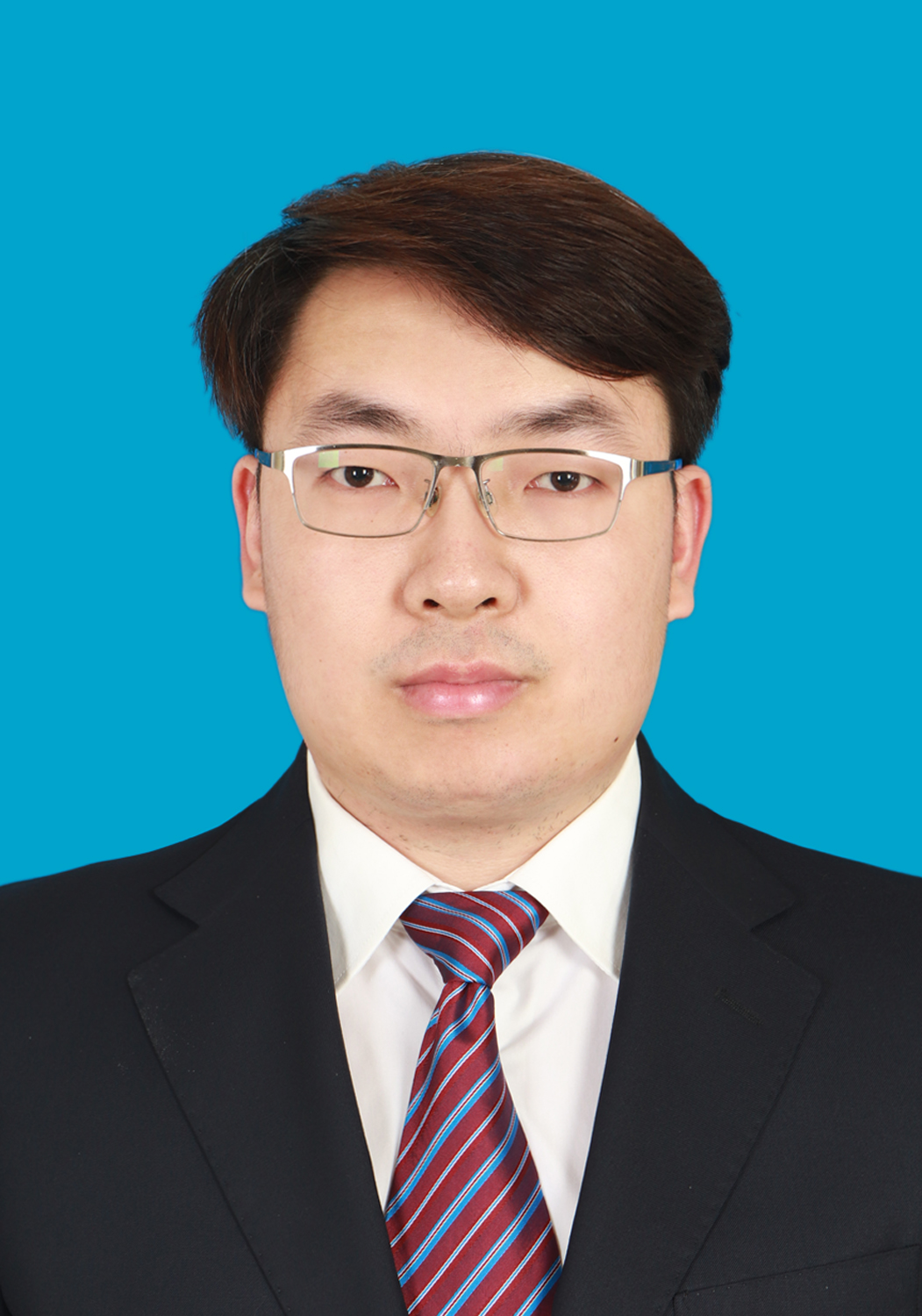}}]{Hao Wu} received his Bachelor's and Master's degrees from the Department of Electronic Engineering, Tsinghua University, in 2017 and 2020, respectively. He is currently pursuing the Ph.D. degree with the Department of Electronic Engineering, Tsinghua University. His major research interests include traffic engineering, network management, and software-defined networking.
\end{IEEEbiography}

\begin{IEEEbiography}[{\includegraphics[width=1in]{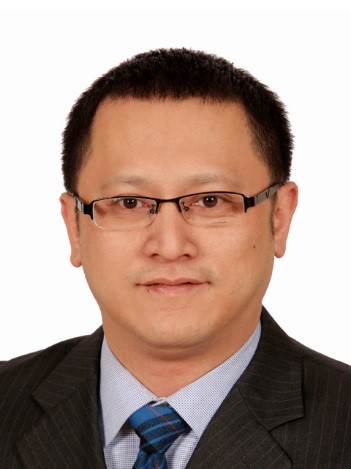}}]{Jian Yan} received his B.S., M.S., and Ph.D. degrees in electronic engineering from Tsinghua University, Beijing, China, in 1998, 2000, and 2010, respectively. He is now a research fellow with the Beijing National Research Center for Information Science and Technology, Tsinghua University. His research interests are mainly in the area of satellite communications. 
\end{IEEEbiography}

\begin{IEEEbiography}[{\includegraphics[width=1in]{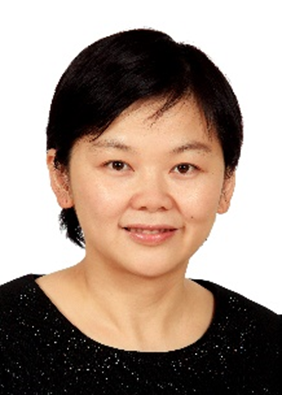}}]{Linling Kuang} received her B.S. and M.S. degrees from the National University of Defense Technology, Changsha, China, in 1995 and 1998, respectively, and her Ph.D. degree in electronic engineering from Tsinghua University, Beijing, China, in 2005. She is now a Research Fellow with the Beijing National Research Center for Information Science and Technology, Tsinghua University. Her research interests include wireless broadband communications, signal processing, and satellite communications.
\end{IEEEbiography}
\end{document}